# Micro-Estimates of Wealth for all Low- and Middle-Income Countries

Guanghua Chi[1†], Han Fang[2], Sourav Chatterjee[2], Joshua E. Blumenstock[1†]

**Affiliations:**

[1] School of Information, University of California, Berkeley; Berkeley, USA.

[2] Facebook, Inc.; Menlo Park, USA.

* Corresponding author. Email: jblumenstock@berkeley.edu

† These authors contributed equally to this work.

**Abstract:** Many critical policy decisions, from strategic investments to the allocation of humanitarian aid, rely on data about the geographic distribution of wealth and poverty. Yet many poverty maps are out of date or exist only at very coarse levels of granularity. Here we develop the first micro-estimates of wealth and poverty that cover the populated surface of all 135 low and middle-income countries (LMICs) at 2.4km resolution. The estimates are built by applying machine learning algorithms to vast and heterogeneous data from satellites, mobile phone networks, topographic maps, as well as aggregated and de-identified connectivity data from Facebook. We train and calibrate the estimates using nationally-representative household survey data from 56 LMICs, then validate their accuracy using four independent sources of household survey data from 18 countries. We also provide confidence intervals for each micro-estimate to facilitate responsible downstream use. These estimates are provided free for public use in the hope that they enable targeted policy response to the COVID-19 pandemic, provide the foundation for new insights into the causes and consequences of economic development and growth, and promote responsible policymaking in support of the Sustainable Development Goals.

Many critical decisions require accurate, quantitative data on the local distribution of wealth and poverty. Governments and non-profit organizations rely on such data to target humanitarian aid and design social protection systems (*1*, *2*); businesses use this information to guide their marketing and investment strategies (*3*); these data also provide the foundation for entire fields of basic and applied social science research (*4*).

Yet reliable socioeconomic data are expensive to collect, and only half of all countries have access to adequate data on poverty (*5*). In some cases, the data that do exist are subject to political capture and censorship (*6*, *7*), and very rarely do such data allow for disaggregation beyond the largest administrative level (*8*). The scarcity of quantitative data is thus a major impediment to policymakers and researchers interested in solutions to global poverty and inequality. Data gaps similarly hinder the broad international coalition working toward the Sustainable Development Goals, in particular toward the first goal of ending poverty in all its forms everywhere (*9*).

To address these data gaps, researchers have developed several approaches to construct poverty maps from non-traditional sources of data. These include methods from small area statistics that combine household sample surveys with comprehensive census data (*10*), and more recent use of satellite 'night-lights' (*11*–*13*), mobile phone data (*14*), social media data (*15*), high-resolution satellite imagery (*16*–*19*), or some combination of these (*20*, *21*). But these efforts have focused on a single continent or a select set of countries, limiting their relevance to development objectives that require a more global perspective.

Here we develop a novel approach to construct micro-regional wealth estimates, and use this method to create the first complete set of micro-estimates of the distribution of poverty and wealth across all 135 LMICs (Fig. 1a). We use this method to generate, for each of roughly 19.1 million unique 2.4km micro-regions in all global LMICs, an estimate of the average absolute wealth (in dollars) and relative wealth (relative to others in the same country) of the people living in that region. These estimates, which are more granular and comprehensive than previous approaches, make it possible to see extremely local variation in wealth disparities (Fig. 1b and Fig. 1c).

Our approach, outlined in Fig. 2, relies on "ground truth" measurements of household wealth collected through traditional face-to-face surveys with 1,457,315 unique households living in 66,819 villages in 56 different LMICs around the world (Table S1). These Demographic and Health Surveys (DHS), which are independently funded by the U.S. Agency for International Development, contain detailed questions about the economic circumstances of each household, and make it possible to compute a standardized indicator of the average asset-based wealth of each village (see SM1) (*8*). We then use spatial markers in the survey data to link each village to a vast array of non-traditional digital data. This includes high-resolution satellite imagery, data from mobile phone networks, topographic maps, as well as aggregated and de-identified connectivity data from Facebook (Table S2). These data are processed using deep learning and other computational algorithms, which convert the raw data to a set of quantitative features of each village (Fig. S2). We use these features to train a supervised machine learning (ML) model that predicts the relative wealth (Fig. 1a) and absolute wealth (Fig. S3a) of each all populated 2.4km grid cells in LMICs (see SM2-4).

The estimates of wealth and poverty are quite accurate. Depending on the method used to evaluate performance, the model explains 56-70% of the actual variation in household-level wealth in LMICs (Fig. 3a). This performance compares favorably to state-of-the-art methods that focus on single countries or continents (*16*, *19*) (see SM4).

To provide visual intuition for the fine granularity of the wealth estimates, Fig. 1c shows an enlargement of a region in the outskirts of Cape Town, South Africa. The satellite imagery shows the physical terrain, which juxtaposes high-density urban areas with farmland and undeveloped zones by the airport and off the main highway. The bottom half of the figure shows the wealth estimates for the same region, which highlight the contrast in wealth between these neighboring areas.

To validate the accuracy of these estimates, and to eliminate the possibility that the ML model is 'overfit' on the DHS surveys, we compare the model's estimates to four independent sources of ground truth data. The first test uses data from 15 LMICs that have collected and published census data since 2001 (Table S3). These data contain census survey responses from 27 million unique individuals, including questions about the economic circumstances of each household. Importantly, the census data are independently collected and are never used to train the ML model. In each country, we aggregate the census data at the smallest administrative unit possible and calculate a 'census wealth index' as the average wealth of households in that census unit. We separately aggregate the 2.4km wealth estimates from the ML model to the same administrative unit. The ML model explains 72% of the variation in household wealth across the 979 census units formed by pooling data from the 15 censuses (Fig. 3c) and, on average, 86% of the variation in household wealth within each of the 15 countries (Fig. S4).

To test the accuracy of the model at the most granular level possible, we obtain three additional sources of survey data that link household wealth information to the exact geocoordinates of each surveyed household. The first dataset, collected by the government of the Togolese Republic (Togo) in 2018-2019, contains a nationally-representative sample of 6,172 households located in 922 unique 2.4km grid cells (Fig. 4a). We find that the ML model's predictions explain 76% of the variation in wealth of these grid cells (Fig. 4b), and 84% of the variation in wealth of cantons, Togo's smallest administrative unit (Fig. 4c). The second dataset, similar to the first but independently collected by the government of Nigeria in 2019, contains a nationally-representative sample of 22,104 households in 2,446 grid cells (Fig. 4d). We find that the ML estimates explain 50% of the variation in grid cell wealth (Figure 4e) and 71% of the variation in wealth of Local Government Areas (Fig. 4f).

We further validate the grid-level predictions using a dataset collected by GiveDirectly, a nonprofit organization that provides humanitarian aid to poor households. In 2018, GiveDirectly surveyed 5,703 households in two counties in Kenya (Fig. 4g), recording a Poverty Probability Index as well as the exact geocoordinates of each household (Fig. 4h). Using these data, we show that even within small rural villages, the ML model's predictions correlate with GiveDirectly's estimates of poverty and wealth (Fig. 4i).

In addition to providing point estimates of the average wealth of the households in each grid cell, we calculate confidence intervals around each estimate (Fig. S3b). These are obtained through standard resampling methods, combined with a more structural approach that models the prediction error as a function of observable characteristics of each location (see SM11). As expected, we find that prediction errors are larger in regions that are far from areas covered by the DHS data (Table S4). While measures of uncertainty are not common in prior work on sub-regional wealth estimation, we believe this is an important step to help promote the responsible use of such estimates in research and policy settings (*22*).

We are making these micro-regional estimates of relative wealth and poverty, along with the associated confidence intervals, freely available for public use and analysis. These estimates are provided through an open and interactive data interface that allows scientists and policymakers to explore and download the data (Fig. S1; see http://beta.povertymaps.net/ for a preliminary "beta" version of the interactive interface).

How might these estimates be used to guide real-world policymaking decisions? One key application is in the targeting of social assistance and humanitarian aid. In the months following the onset of the COVID-19 pandemic, hundreds of new social protection programs were launched in LMICs, and in each case, program administrators faced difficult decisions about whom to prioritize for assistance (*23*). This is because in many LMICs, planners do not have comprehensive data on the income or consumption of individual households (*24*). The new estimates provide one potential solution.

In simulations, we find that geographic targeting using our micro-estimates allocates a higher share of benefits to the poor (and a lower share of benefits to the non-poor) than geographic targeting approaches based on recent nationally representative household survey data (Table 1 and SM13). This is because the micro-estimates make it possible to target smaller geographic regions than would be possible with traditional survey data – a finding that is consistent with prior work that suggests that more granular targeting can produce large gains in welfare (*2*, *25*, *26*). For instance, the most recent DHS survey in Nigeria only surveyed households in 13.8% of all Nigerian wards (the smallest administrative unit in the country); by contrast, the micro-estimates cover 100% of wards. In Togo, existing government surveys only provide poverty estimates that are representative at the regional level (of which there are only 5); we provide estimates for 9,770 distinct tiles.

Based on the strength of these results, the Government of Nigeria is using these estimates as the basis for social protection programs that are providing benefits to millions of poor families (*27*). Likewise, the Government of Togo is using these estimates to target mobile money transfers to hundreds of thousands of the country's poorest mobile subscribers (*28*). These examples highlight how the ML estimates can improve targeting performance even in countries with robust national statistical offices, like Nigeria and Togo. In the large number of LMICs that have not conducted a recent nationally representative household survey, these micro-estimates create an option for geographic targeting that would otherwise not exist.

The standardized procedure through which these estimates are produced may also be attractive in contexts where political economy considerations might lead to systematic misreporting of data (*7*) or influence whether new data are collected at all (*6*). However, this does not imply the ML estimates are apolitical, as maps have a historical tendency to perpetuate existing relations of power (*29*). One particular concern is that the technology used to construct these estimates may not be transparent to the average user; if not produced or validated by independent bodies, such opacity might create alternative mechanisms for manipulation and misreporting.

While our primary focus is on constructing, validating, and disseminating this new resource, the process of building this dataset produces several insights relevant to the construction of high-resolution poverty maps. For instance, we find that different sources of input data complement each other in improving predictive performance (*20, 21*). While prior work has focused heavily on satellite imagery, we find that models trained only on satellite data do not perform as well as models that include other input data (Fig. S7a). In particular, information on mobile connectivity is highly predictive of sub-regional wealth, with 5 of the 10 most important features in the model related to connectivity (Fig. S2 and SM5).

The global scale of our analysis also reveals intuitive patterns in the geographic generalizability of machine learning models (*16, 30, 31*). We find that models trained using data in one country are most accurate when applied to neighboring countries (Fig. S6). Models also perform better in countries when trained on countries with similar observable characteristics (Table S4). And while much of the model's performance derives from being able to differentiate between urban and rural areas, the model can differentiate variation in wealth within these regions as well (Fig. 3b).

Our hope is that these methods and maps can provide a new set of tools to study economic development and growth, guide interventions, monitor and evaluate policies, and track the elimination of poverty worldwide.

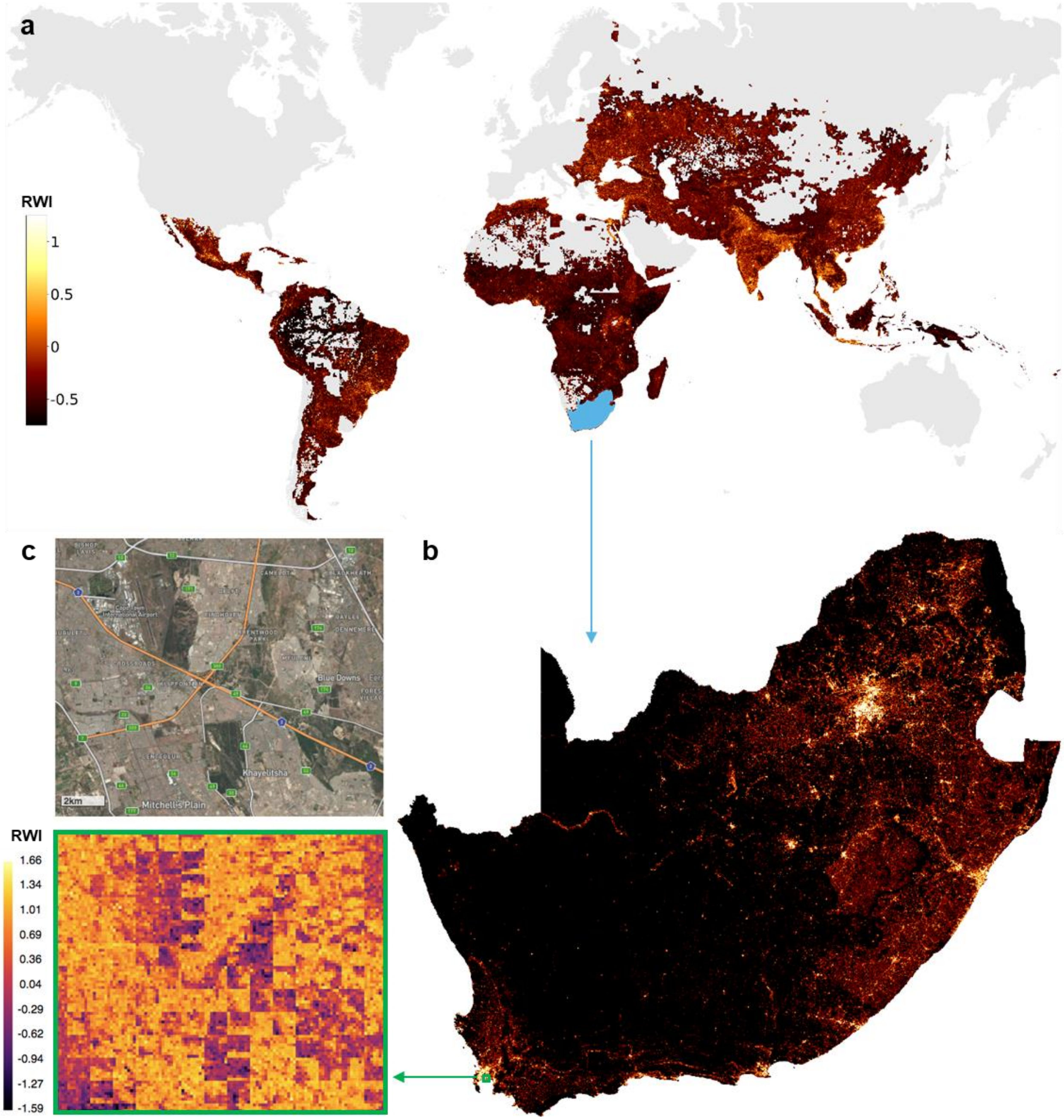

**Fig. 1 | Micro-estimates of wealth for all low- and middle-income countries**. **a)** Estimates of the relative wealth of each populated 2.4km gridded region of all 135 LMICs. Interactive version available at http://beta.povertymaps.net/. Enlargements show **b)** the countries of South Africa and Lesotho; **c)** The 12x12km region around the Khayelitsha township near Cape Town, with 0.58m-resolution estimates (both panels show the same region).

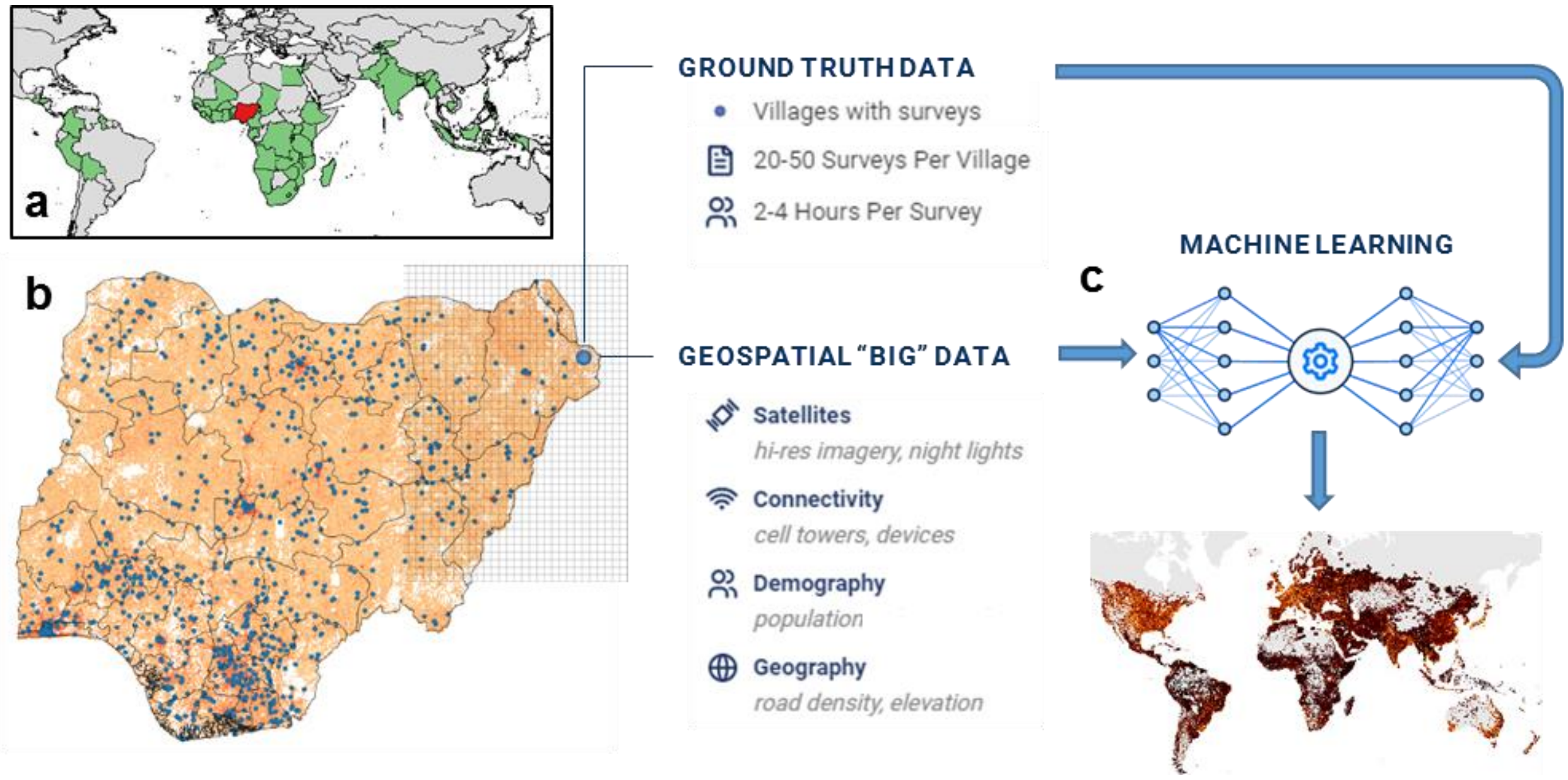

**Fig. 2 | Overview of approach**. **a)** Nationally-representative household survey data is obtained from 56 different countries around the world. **b)** In Nigeria, for example, there are 40,680 households surveyed in 899 unique survey locations ("villages"). Non-traditional data from satellites and other existing sensors are also sourced from each location. **c)** These data are used to train a machine learning algorithm that predicts micro-regional poverty from non-traditional data, even in regions where no ground truth data exists.

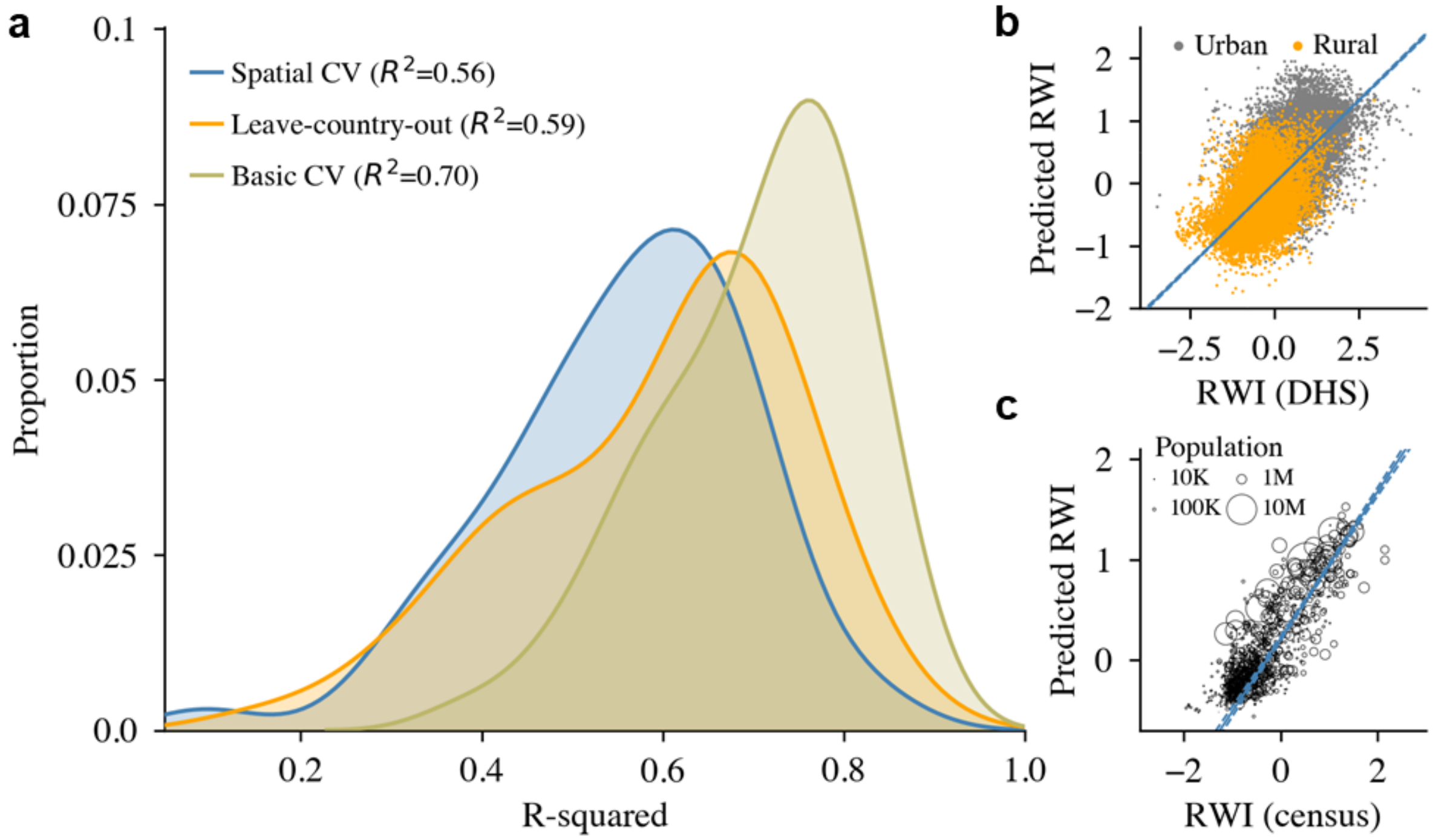

**Fig. 3 | Model performance**. **a)** Distribution of model performance, across 56 countries with ground truth data, using 3 different approaches to cross-validation. **b)** Much of the model's predictive power comes from being able to differentiate between rural and urban locations, but the model also detects wealth differentials within urban and rural locations. **c)** The ML model explains 72% of the variation in wealth, as measured with independent census data from 15 LMIC's. Population-weighted regression lines in blue; 95% confidence intervals in dashes.

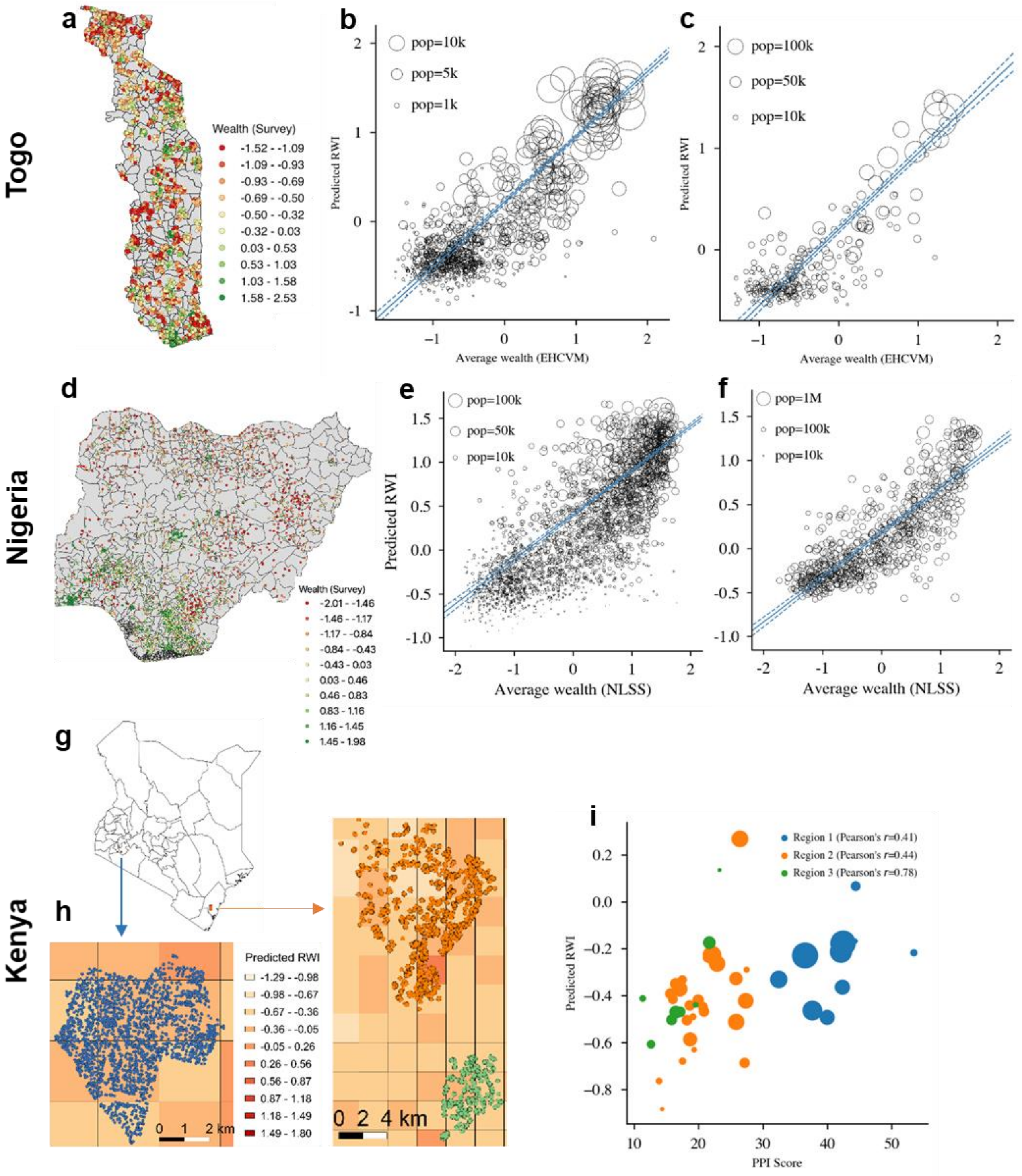

**Fig. 4 | Validation with independently collected micro-data in Togo, Nigeria, and Kenya**. **a)** Map of Togo showing locations of surveyed households (jitter added to map to preserve household privacy). **b)** Scatterplot of the predicted RWI of each grid cell (y-axis) against the average wealth of the grid cell, as reported in a nationally-representative government survey. Points sized by population. Population-weighted regression lines in blue; 95% confidence intervals in shown with dashes. **c)** Scatterplot of predicted RWI against average wealth of each

canton, the smallest administrative unit in the country. **d)** Map of surveyed households in Nigeria (jitter added). **e)** Scatterplot of predicted RWI against average wealth of each grid cell. **f)** Scatterplot of predicted RWI against average wealth of each local government area (LGA). **g)** Map of Kenya showing the regions surveyed by GiveDirectly. **h)** Enlargement of the three survey regions, showing the location of each of 5,703 surveyed households. Colors of background grid cells indicates RWI predicted from the ML model. In both enlargements, the width of the grid cell is 2.4km. **i)** Scatterplot of the predicted RWI of each grid cell (y-axis) against the average PPI of all surveyed households in the grid cell (x-axis). Points are sized by the number of households in the grid cell, and colored by region.

| a) | (1)<br># of spatial units | (2)<br># of units with estimates | (3)<br>$R^2$ | (4)<br>Targeting accuracy, poorest 25% | (5)<br>Targeting accuracy, poorest 50% |
|---|---|---|---|---|---|
| **TOGO** | | | | | |
| *Panel A (Togo): High-resolution estimates* | | | | | |
| Tiles | 10,187 | 10,187 | 0.60 | 0.73 | 0.79 |
| Canton targeting | 387 | 387 | 0.56 | 0.73 | 0.77 |
| *Panel B (Togo): Imputation based on DHS data* | | | | | |
| Prefecture targeting | 40 | 40 | 0.49 | 0.70 | 0.70 |
| Canton targeting | 387 | 185 | 0.52 | 0.76 | 0.80 |
| **NIGERIA** | | | | | |
| *Panel C (Nigeria): High-resolution estimates* | | | | | |
| Tile targeting | 159,147 | 159,147 | 0.53 | 0.79 | 0.79 |
| Ward targeting | 8,808 | 8,808 | 0.51 | 0.78 | 0.78 |
| *Panel D (Nigeria): Imputation based on DHS data* | | | | | |
| State targeting | 37 | 37 | 0.37 | 0.75 | 0.74 |
| LGA targeting | 774 | 631 | 0.47 | 0.78 | 0.76 |
| Ward targeting | 8,808 | 1,218 | 0.54 | 0.83 | 0.79 |

b)

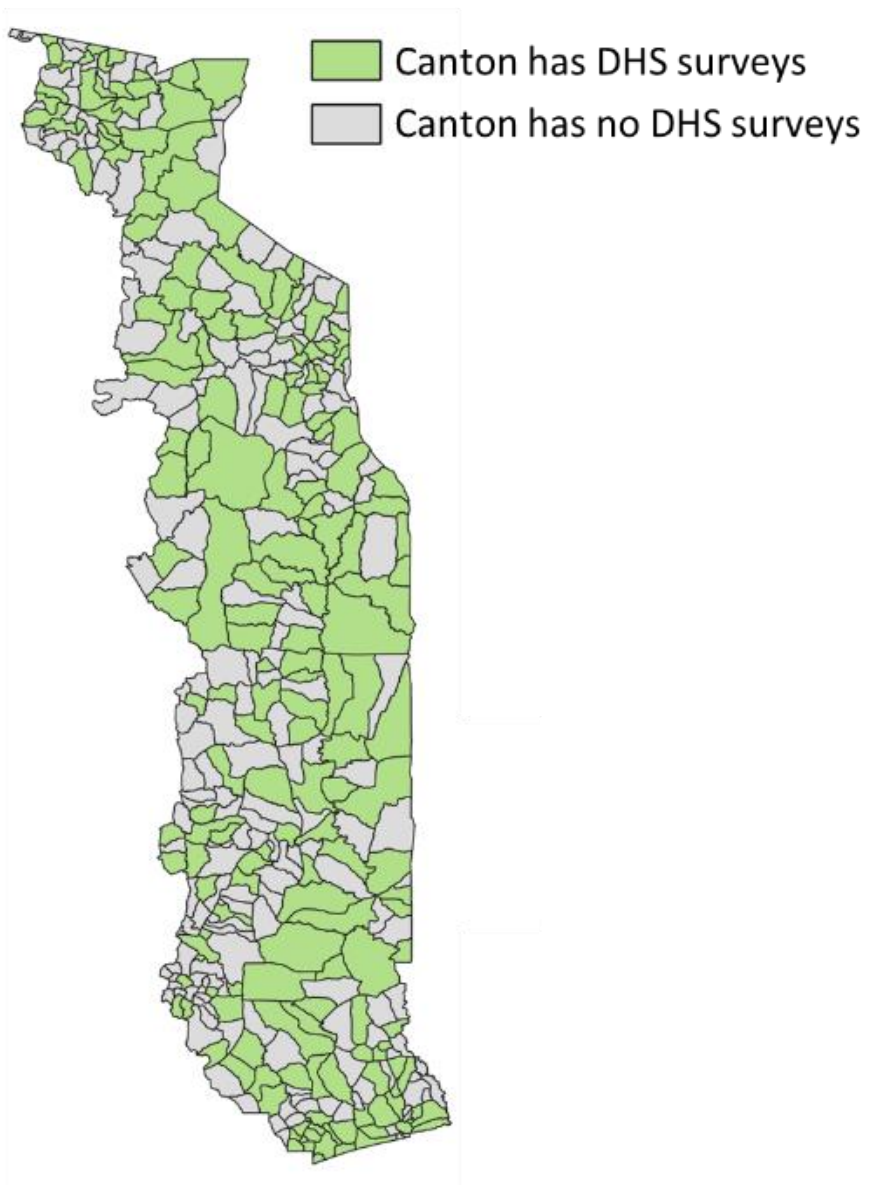

c)

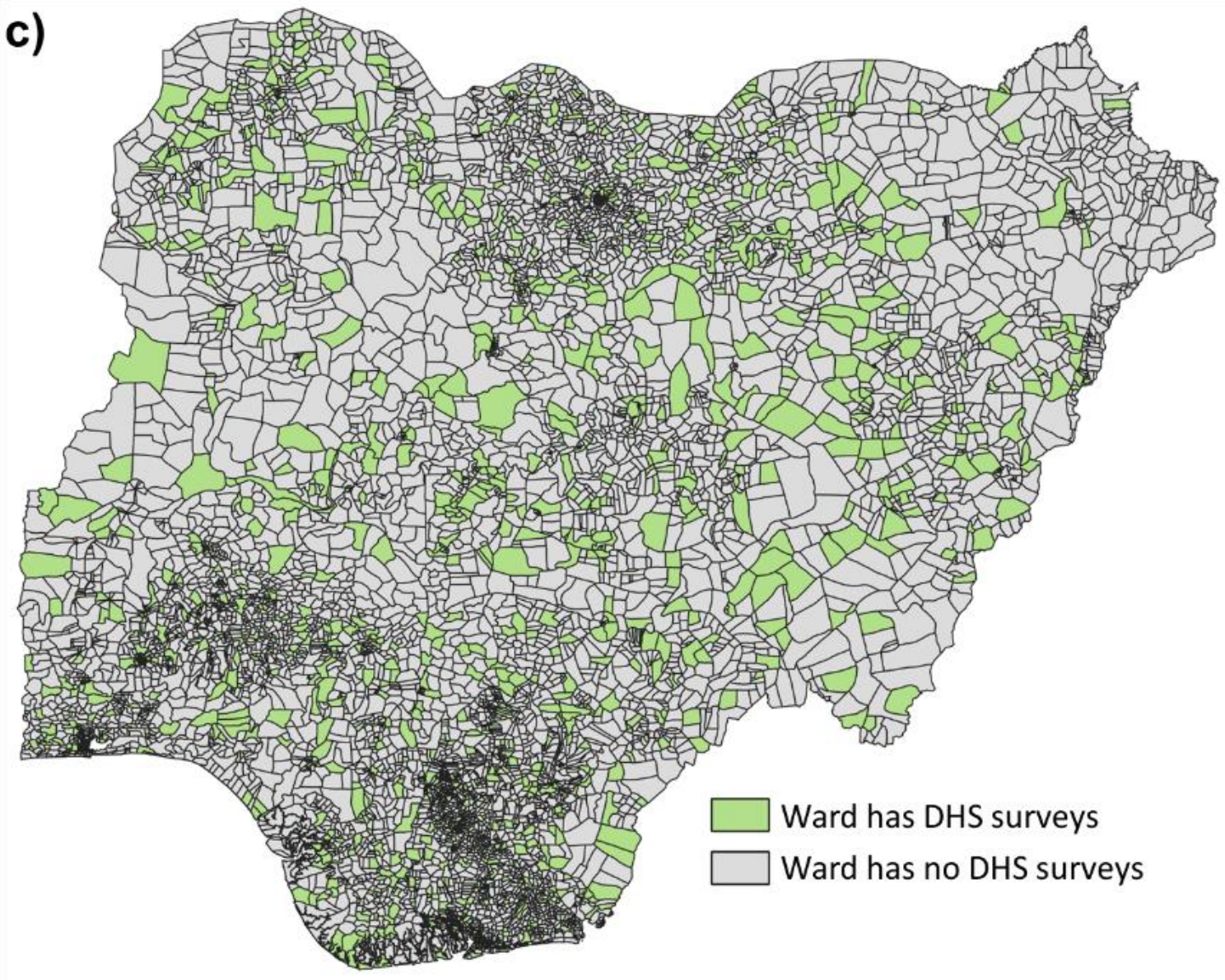

**Table 1 | Targeting simulations in Togo and Nigeria. a)** Panels A and C simulate the performance of anti-poverty programs that geographically targets households using the ML estimates of tile wealth, under scenarios where the program is implemented at the tile level (first row) or smallest administrative unit in the country (second row). Panels B and D simulate the geographic targeting based on the most recent DHS survey, using administrative units of different sizes. When an admin unit has no surveyed households, the wealth of the unit is imputed based on the wealth of the geographic unit closest to the household. See Methods and Table S7 and Table S8 for details. **b)** Map of Togo shows the 47.8% of cantons in Togo in which DHS household surveys were conducted; unsurveyed cantons shown in grey. **c)** Map of Nigeria shows surveyed wards in green (13.83% of wards) and unsurveyed wards in grey (86.17%).

# Supplementary Materials for

## Micro-Estimates of Wealth for all Low- and Middle-Income Countries

Guanghua Chi, Han Fang, Sourav Chatterjee, Joshua E. Blumenstock

Correspondence to: jblumenstock@berkeley.edu

## Materials and Methods

### SM1. Ground truth wealth measurements

The ground truth wealth data used to train the predictive models are derived from household surveys conducted by the Demographic and Health Survey (DHS) Program. According to the program, the DHS collects "nationally-representative household surveys that provide data for a wide range of monitoring and impact evaluation indicators in the areas of population, health, and nutrition."[i] We elected to train our model exclusively on DHS data because it is the most comprehensive single source of publicly available, internationally standardized wealth data that provides household-level wealth estimates with sub-regional geo-markers.

The fact that we use the DHS data as our ground truth measure of wealth and poverty means that we are effectively training our machine learning algorithm to reconstruct a DHS-style relative wealth index – albeit at a much finer spatial resolution and in areas where DHS surveys did not occur. This is because we believe the DHS version of a relative wealth index is the best publicly available instrument for consistently measuring wealth across a large number of LMICs. However, it posits a specific, asset-based definition of wealth that does not necessarily capture a broader notion of human development. More broadly, a rich social science literature debates the appropriateness of different measures of human welfare and well-being (*4*, *33*). Our decision to focus on estimating asset-based wealth, rather than a different measure of socioeconomic status (SES), was motivated by several considerations. First, in developing economies, where large portions of the population do not earn formal wages, measures of income are notoriously unreliable. Instead, researchers and policymakers rely on asset-based wealth indices or measures of consumption expenditures. Between these two, wealth is much less time-consuming to record in a survey; as a result, wealth data are more commonly collected in a standardized format for a large number of countries (*34*).

We obtain the most recent publicly-available DHS survey data from 56 countries (Table S1). The criteria for inclusion are that the data are available for download through the DHS website (as of March 2020), the data contain asset/wealth information and sub-regional geomarkers, and that the most recent survey was conducted since 2000. The combined dataset contains the survey responses from 1,457,315 household surveys taken across Africa, Asia, Europe, and Latin America. Each individual household survey lasts several hours, and contains several questions related to the socioeconomic status of the household. We focus on a standardized set of questions about assets and housing characteristics.[ii] From the responses to these questions, and following standard practice (*8, 35*), the DHS calculates a single continuous measure of relative household wealth, the Relative Wealth Index (RWI), by taking the first principal component of these 15 questions. It is this DHS-computed RWI that we rely upon as a ground truth measure of wealth.

[i] http://www.dhsprogram.com/

In addition to providing measures of wealth for each household, the DHS indicates the *cluster* in which each household is located. The 1.5M households are associated with 66,819 unique clusters, where a cluster is roughly equivalent to a village in rural areas and a neighborhood in urban areas. We calculate the average wealth of each "village" cluster by taking the mean RWI of all surveyed households in that cluster.[iii] This village-level average RWI is the target variable for the machine learning model.

## SM2. Input data

The prediction algorithms rely on data from several different sources (Table S2). To facilitate downstream analysis, all data are converted into *features* that are aggregated at the level of a 2.4km grid cell. We use 2.4km cells because that is the highest resolution at which many of our input data are available, and it best suited to the spatial merge with the survey data (see "supervised machine learning" below). We were also concerned that providing estimates of wealth at even smaller grid cells might compromise the privacy of individual households. Thus, if the native resolution of a data source is higher than 2.4km, we aggregate the smaller cells to the 2.4km level by taking the average of the smaller cells.

The features input into the model indicate, for each cell, properties such as the average road density, the average elevation, and the average annual precipitation. Several features related to telecommunications connectivity are obtained from Facebook, which uses proprietary methods to estimate the availability and use of telecommunications infrastructure from de-identified Facebook usage data[iv]. All estimates are regionally aggregated at the 2.4km level to preserve user privacy. We use estimates of the number of mobile cellular towers in each grid cell, as well as the number of WiFi access points and the number of mobile devices of different types. These measures are based on the infrastructure used by Facebook users, so may not be representative of the full population. To the extent that these features are predictive of regional wealth (which they are), no deeper inference or causal interpretation should be drawn from the empirical association. Rather, these patterns simply indicate that the regional distribution of wealth is correlated with these non-representative measures of telecommunications use.

Since the raw satellite imagery is extremely high-dimensional, we use unsupervised learning algorithms to compress the raw data into a set of 100 features. Specifically, following Jean *et al.* (*16*), we use a pre-trained, 50-layer convolutional neural network to convert each 256x256 pixel image into 2048 features, and then extract the first 100 principal components of these 2048-

---

[ii] The full set of indicators are: electricity in household, telephone, automobile, motorcycle, refrigerator, TV, Radio, water supply, cooking fuel, trash disposal, toilet, floor material, wall material, roof material, and rooms in house.

[iii] Our main estimates do not use the cluster weights provided by the DHS. We separately evaluate a model that used these weights to train a weighted regression tree, and find that the predictions of the two models are highly correlated ($r = 0.9$), and result in similar overall performance ($R^2$=0.56 without weights vs. $R^2$=0.54 with weights).

[iv] https://research.fb.com/category/connectivity/

dimensional vectors.[v] These 100 components explain 97% of the variance of the 2048 features (Fig. S8).

All input features are normalized by subtracting the country-specific mean and dividing by the country-specific standard deviation.

### SM3. Spatial join

We match the ground truth wealth data to the input data using spatial information present in both datasets. The 2.4km grid cells are defined by absolute latitude and longitude coordinates specified by the Bing tile system.[vi] The DHS data include approximate information about the GPS coordinate of the *centroid* of each of the 66,819 villages. However, the exact geocoordinates are masked by the DHS program with up to 2km of jitter in urban areas and up to 5km of jitter in rural areas.

To ensure that the input data associated with each village cover the village's true location, we include a 2x2 grid of 2.4km cells around the centroid in urban areas, and a 4x4 grid in rural areas. For each of village, we then take the population-weighted average of the 112-dimensional feature vectors across 2x2 or 4x4 set of cells, using existing estimates of the population of 2.4km grid cells (*37*). This leaves us with a training set of 66,819 villages with wealth labels (calculated from the ground truth data) and 112-dimensional feature vectors (computed from the input data).

### SM4. Supervised machine learning

We use machine learning algorithms to predict the average RWI of each village from the 112 features associated with that village. We do not perform ex ante feature selection prior to fitting the model. We use a gradient boosted regression tree, a popular and flexible supervised learning algorithm, to map the inputs to the response variable. To tune the hyperparameters of the gradient boosted tree, we use three different approaches to cross-validation.[vii]

- *K-fold cross-validation* (labeled "Basic CV" in Fig. 3a). For each country, the labelled data are pooled, and then randomly partitioned into $k = 5$ equal subsets. A model is trained on all but one subset and tested on the held-out subset. The process is repeated $k$ times and we report average held-out performance for that country. This approach to cross-validation is used most frequently in prior work, but can substantially over-estimate performance(*38*).

---

[v] We use a 50-layer resnet50 network (*36*), where pre-training is similar to Mahajan et. al. (*32*). This network is trained on 3.5 billion public Instagram images (several orders of magnitude larger than the original Imagenet dataset) to predict corresponding hashstags. We extract the 2048-dimensional vector from the penultimate layer of the pre-trained network, without fine-tuning the network weights. The satellite imagery has a native resolution of 0.58 meters/pixel. We downsample these images to 9.375m/pixel resolution by averaging each 16x16 block. The downsampled images are segmented into 2.4km squares, then passed through the neural network. For each satellite image, we do a forward-pass through the network to extract the 2048 nodes on the second-to-last layer. We then apply PCA to this 2048-dimensional object and extract the first 100 components. The PCA eigenvectors are computed from images in the training dataset (i.e., the images from the 56 countries with household surveys).

[vi] See https://docs.microsoft.com/en-us/bingmaps/articles/bing-maps-tile-system

[vii] Hyperparameters were tuned to minimize the cross-validated mean squared error, using a grid search over several possible values for maximum tree depth (1, 3, 5, 10, 15, 20, 30) and the minimum sum of instance weight needed in a child (1, 3, 5, 7, 10).

This bias arises because both the input (e.g., satellite) and response (RWI) data are spatially auto-correlated, leaving the training and test data not i.i.d. (*39*).[viii]

- *Leave-one-country-out cross-validation* ("Leave-country-out"). For each country, a model is trained using the pooled data from all other 55 countries; the test performance is evaluated on the held-out country (*16*).
- *Spatially-stratified cross-validation* ("Spatial CV"). This method ensures that training and test data are sampled from geographically distinct regions (*38, 39*). In each country, we select a random cell as the training centroid, then define the training dataset as the nearest ($k$-1)/$k$ percent of cells to that centroid. The remaining 1/$k$ cells from that country form the test dataset. This procedure is repeated $k$ times in each country.

Fig. 3a compares the performance of these three methods, by showing the distribution of $R^2$ values for each approach to cross-validation (the distribution is formed from 56 countries, where a separate model is trained and cross-validated in each country). The difference in $R^2$ resulting from different approaches to cross-validation highlights the potential upward bias in performance that results from spatial auto-correlation in training and test data. By comparison, recent work on wealth prediction in Africa found that a mixture of remote sensing and nightlight imagery explains on average 67% of the variation in wealth (*19*). That benchmark was based on an approach similar to the "leave-country-out" method shown in Fig. 3a; the slight decline in performance that we observe is likely due to the fact that the 23 countries in Africa studied by (*19*) are substantially more homogenous than the full set of LMICs that we analyze.

Unless noted otherwise, all analysis in this paper uses models based on spatially-stratified cross-validation. While this has the effect of lowering the $R^2$ values that we report, we believe it is the most conservative and appropriate method for training machine learning models on geographic data with spatial auto-correlation.

## SM5. Feature importance

To shed light on which of the various data sources are driving the model's predictions, Fig. S2 provides two different indicators of feature importance. Fig. S2a (left panel) indicates the unconditional correlation between the true wealth label and each individual feature, calculated as the $R^2$ from a univariate regression of the wealth label on each single feature (each row is a separate regression; with 56 countries, there are 56 $R^2$ values that form the distribution of each boxplot). Fig. S2b (right panel) indicates the model gain, which provides an indication of the relative contribution of each feature to the final model (specifically, it is the average gain across all splits in the random forest that use that feature)(*40*). In general, we find that data related to connectivity, such as the number of cell towers and mobile devices in a region, are the most predictive features; nightlight radiance and population density are also predictive. While no single feature derived from satellite imagery is especially predictive in isolation, the large number of satellite features collectively contribute to model accuracy – this can be seen most

[viii] In an extreme example, imagine a single town that covers two adjacent grid cells. If one of the grid cells is in the training set and the other is in the testing set, a flexible model could simply learn to detect the town and predict its wealth. This sort of overfitting is not addressed by standard $k$-fold cross-validation.

directly in Fig. S7a, which compares the predictive performance of models with and without satellite imagery.

### SM6. Out-of-sample estimates

To produce the final maps and micro-estimates, as well as the public dataset, we pool data from all 56 countries and train a single model using spatially-stratified cross-validation to tune the model parameters.[ix] This model maps 112-dimensional feature vectors to wealth estimates. We then pass the 112-dimensional feature vector for each 2.4km grid cell located in a LMIC through this trained model to produce an estimate of the relative wealth (RWI) of each grid cell (Fig. 1). We use the World Bank's List of Country and Lending Groups to define the set of 135 low- and middle-income countries.[x] Since we do not normalize these predictions at the country level after they have been generated, we do not expect that each country will have the same within-country RWI distribution (i.e., the amount of "bright" and "dark" spots will differ between countries).

To help preserve the privacy of individuals and households, we do not display wealth estimates for 2.4km regions where existing population layers indicate the presence of 50 or fewer individuals in the region (*37*). Instead, we aggregate neighboring 2.4km tiles (by taking the population-weighted average RWI) until the total estimated population of the larger area is at least 50. The "neighbors" of a tile are those tiles that fall within the larger tile, using the tile boundaries defined by the Bing tile system.[xi] All of the neighboring 2.4km cells in the larger tile are then assigned the same estimate of RWI (i.e., the population-weighted average).

### SM7. Cross-sectional estimation

Our main objective is to produce accurate estimates of the current, cross-sectional distribution of wealth and poverty within LMICs. In training the machine learning model described above, we thus use the most recently available version of each data source. The ground truth wealth measurements cover a wide range of years (Table S1); the input data are primarily generated in 2018 (Table S2). This often creates a mismatch between the dates of the input variables and the survey labels for a given region. In practice, this means that our estimates are best at capturing within-country variation in wealth that does not change over a relatively short time horizon (i.e., between the prior survey date and 2018). Analysis of DHS data from LMICs with multiple surveys suggest a high degree of persistence in the within-country variation in wealth (Fig. S11)[xii]. Still, this approximation likely introduces error into our model, and suggests that these

---

[ix] In robustness analysis, we separately constructed complete micro-estimates for all LMICs in which the estimates for all countries *without* DHS surveys were based on the full model trained on pooled data from the 56 countries with DHS surveys; then, in each of the 56 countries *with* DHS surveys, we replaced the pooled estimates with the estimates from a model trained exclusively with data from the target country. We find that the average accuracy of this alternative approach ($R^2 = 0.54$, using spatial CV) is nearly identical to the pooled approach (average $R^2 = 0.56$, using spatial CV).

[x] We use the 2018 version of this list, which includes countries whose Gross National Income per capita was less than $4,045. See https://datahelpdesk.worldbank.org/knowledgebase/articles/906519-world-bank-country-and-lending-groups

[xi] The 2.4km estimates correspond to Bing tile level 14; the next largest tile, Bing tile level 13, defines 4.8km grid cells, and so forth.

[xii] Across the 33 countries with two or more DHS surveys conducted since 2000, the median $R^2$ between regional (admin-2) wealth estimates from the most recent DHS survey and the preceding DHS survey is 0.81.

estimates are better suited toward applications that require a measure of permanent income than to applications that require an understanding of poverty dynamics. More broadly, see this model's performance as a benchmark that can be improved upon as more input and survey data become available.

In an ideal world, we would obtain historical input data from the same years in which each survey was conducted. Unfortunately, historical versions of most of the input data described in Table S2 do not exist. Alternatively, we could restrict our analysis to input data that do exist in a historical panel. However, as shown in Fig. S7a, excluding key predictors substantially limits the model's predictive accuracy. Another option would be to only train the model using more recent surveys. In Fig. S9a, we observe that the accuracy of a model trained on the subset of 24 countries that conducted DHS surveys since 2015 is quite similar to the performance of a model trained on all 56 countries with DHS data since 2000. Related, when we validate the model's performance using independently collected census data (see below for details), we find no evidence to suggest that a shorter gap between the date of the DHS training data and the data of the census increases the predictive accuracy of the model (Fig. S10).

## SM8. Independent validation with census data

We validate the accuracy of the ML estimates using census data that are collected independently from the DHS data used to train the models. Specifically, we obtain census data from all countries with public IPUMS-I data, where the census occurred since 2000 and where asset data are complete (*41*). In total, these data cover 15 countries on 3 continents, and capture the survey responses of 27 million individuals (Table S3). We assign each of these individuals a census wealth index by taking the first principal component of the 13 assets present in the census data. This list is similar to the DHS asset list, but excludes data on motorcycles and rooms in the household. As with the DHS data, the PCA eigenvectors are computed separately for each country. Finally, we compute the average census wealth index over all households within each second administrative unit, the smallest unit that is consistently available across countries. Of the 1,003 census units, 979 have households with wealth information and also contain a 2.4km tile with a centroid inside the unit.

Fig. 3c shows a scatterplot of these 979 administrative units, sized by population. The x-axis indicates the average wealth of each administrative unit, according to the census (calculated as the mean first principal component across all households in the unit). The y-axis indicates the average predicted RWI of the administrative unit, calculated by taking the population-weighted mean RWI of all grid cells within the unit. The population-weighted regression line $R^2 = 0.72$ (obtained when pooling the 979 admin-2 regions from all 15 countries). Fig. S4 disaggregates Fig. 3c by country, showing the relationship between census-based wealth and RWI across the administrative units of each country. The average population-weighted $R^2$ across the 15 countries is 0.86 (Table S3);

## SM9. High-resolution validation with independently collected micro-data from Togo, Nigeria, and Kenya

We further validate the accuracy of the ML estimates at the finest possible spatial resolution by comparing them to three independently-collected household surveys in Togo, Nigeria, and Kenya. In each case, we obtain the original survey data for all households, as well as the exact GPS coordinates of each surveyed household. As with the census data, none of these datasets were used to train the ML model; they thus provide an independent and objective assessment of the accuracy and validity of our new estimates.

*Togo*. As part of the 2018-2019 Enquete Harmonisee sur les Conditions de Vie des Menages (EHCVM), the government of Togo conducted a nationally-representative household survey with 6,172 households.[xiii] A key advantage of these data is that, in addition to observing a wealth index for each household (calculated as the first principal component of roughly 20 asset-related questions), we observe each household's exact geo-coordinates (Fig. 4a). The 6,172 households are located in 922 unique 2.4km grid cells (which correspond to 260 unique cantons, the smallest administrative unit in Togo), of the 9,770 total grid cells in the country. We also note that there is nothing Togo-specific in how the ML model is trained: we simply use the estimates generated by the final model that is trained using spatially-stratified cross-validation from all 56 countries with DHS data (also shown in Fig. 1).

Fig. 4a shows the approximate location of each of the households surveyed in the EHCVM. Fig. 4b compares, for each of the 922 grid cells with surveyed households, the average wealth of all households in each grid cell as calculated from the EHCVM (x-axis) to the estimated RWI of the grid cell, which is displayed on the y-axis ($R^2 = 0.76$). Fig. 4c presents an analogous analysis for each of the 260 cantons in Togo, where the x-axis indicates the average EHCVM wealth of all households in the canton and the y-axis indicates the average RWI for each canton, calculated as population-weighted mean of all cells within the canton ($R^2 = 0.84$).

*Nigeria*. During the 2018-2019 Nigerian Living Standards Survey (NLSS), Nigeria's National Bureau of Statistics, in collaboration with the World Bank, conducted a nationally-representative household survey with 22,104 households (Fig. 4d).[xiv] Like the EHCVM in Togo, the NLSS in Nigeria contains a wealth index for each household and each household's exact geo-coordinates. The 22,104 households are located in 2,446 unique 2.4km grid cells. We compare the NLSS micro-data, which were never used to train the model, to the final estimates of the ML.

Fig. 4d shows the approximate location of each of the households surveyed in the NLSS. Fig. 4e compares, for each of the 2,446 grid cells with surveyed households, the average wealth of all households in the grid cell as calculated from the NLSS to the estimated RWI of the grid cell ($R^2 = 0.50$). Fig. 4f presents an analogous analysis for each of the 774 Local Government Areas (LGAs) in Nigeria ($R^2 = 0.71$).

*Kenya*. We also validate the accuracy of the grid-cell RWI estimates using GPS-enabled survey data collected in the Kenyan counties of Kilifi and Bomet (Fig. 4g). These data were collected by

---

[xiii] See https://inseed.tg/

[xiv] Borno State was excluded for security concerns. See https://www.nigerianstat.gov.ng/nada/index.php/catalog/64

GiveDirectly, a nonprofit organization that provides unconditional cash transfers to poor households in East Africa.[xv] When GiveDirectly works in a village, they conduct a socioeconomic survey with every household in the village. The survey includes a standardized set of 10 questions that form the basis for a Poverty Probability Index (PPI)[xvi], which GiveDirectly uses to determine which households are eligible to receive cash transfers. GiveDirectly also records the exact geocoordinates of each household that they survey (Fig. 4g).

Fig. 4i compares estimates of micro-regional wealth based on GiveDirectly's household PPI census to corresponding estimates of wealth based on the ML model. We calculate the average PPI score of each 2.4km grid cell by taking the mean of the PPI scores of all households in the grid cell. We compare this to the predicted RWI from the ML model. Across the 44 grid cells shown in Fig. 4h (10 from region 1; 26 from region 2; 8 from region 3), the predicted RWI explains 21% of the variation in PPI (Pearson's $r = 0.46$). Within each region, the correlation between PPI and RWI ranges from 0.41 – 0.78.

While the ML model explains less of the variation in Kenya than it does in Togo, Nigeria, or in the 15 census countries, this is a much more stringent test. This is because the comparison is being done across 44 spatially proximate units (Figure 4h) in 3 small and relatively homogenous villages. Within these villages, there is less variation in wealth than there is across an entire country (the variance in RWI across the 44 cells is 0.05; across all of Kenya the variance is 0.10). Our other tests - and indeed all prior work of which we are aware – measure $R^2$ across entire countries. The Kenya test is also handicapped by the fact that the Kenyan PPI is not strictly a wealth index, containing questions about education, consumption, and housing materials. Measures of wealth and poverty are quite sensitive to the measurement instrument used.[xvii] To our knowledge, this is the first attempt to compare estimates of micro-regional wealth, based on variation within single villages, to independently-collected household survey data where the exact location of each surveyed household is known. We therefore find it encouraging that the predicted RWI roughly separates wealthier from poorer neighborhoods within these small regions.

### SM10. Model accuracy in high-income nations

The primary intent of the model is to produce estimates of wealth in LMICs, and it is from LMICs that we source all of the ground truth data used to train the model. For completeness, we assess the performance of the model's predictions in high-income nations. This comparison is imperfect, because high-income nations do not typically collect asset-based wealth indices, which is what the ML model is trained to estimate. Instead, we compare the Absolute Wealth Estimates (AWE) of the ML model (see below for details on how these are constructed) to independently-produced data on regional Gross Domestic Product per capita (GDPpc) from 30 member nations of the Organisation for Economic Co-operation and Development (OECD).

---

[xv] http://www.givedirectly.org/

[xvi] See https://www.povertyindex.org/country/kenya

[xvii] For instance, Filmer and Pritchett find that, even within a single survey, the Spearman rank correlation between an asset index and a measure of consumption expenditures ranges from 0.43 (in Pakistan) to 0.64 (in Nepal).[21]

These data are collected by the National Statistical Offices of each respective country, through the network of Delegates participating in the Working Party on Territorial Indicators.[xviii]

In each country, we obtain the OECD's estimate of the average GDPpc of each 'small' (TL3) region.[xix] We separately calculate the AWE of each region by taking the population-weighted average AWE of all 2.4km grid cells in the region. Fig. S5a shows a scatterplot of these 1540 administrative units, sized by population, where the x-axis indicates the OECD-based measure of wealth of the administrative unit and the y-axis indicates the population-weighted average predicted AWE of the administrative unit. Fig. S5b shows the accuracy of the model in each of the 30 countries. The average population-weighted $R^2$ across the 30 countries is 0.50; the population-weighted regression line $R^2 = 0.59$ (obtained when pooling the 1540 regions from all 30 countries). We note that the AWE values are generally larger than the OECD estimates of GDPpc (the slope of the regression line in Fig. S5a is 1.35). This is likely due to the fact that the GDPpc estimates used to construct AWE (sourced from the World Bank) are consistently higher than the GDPpc estimates sourced from the OECD. This comparison is made in Fig. S5c, where we compare, for the 30 OECD nations, the relationship between the World Bank estimate of GDPpc and the average regional GDPpc based on OECD data (the slope of the regression line Fig. S5c is 1.66).

## SM11. Confidence intervals and model error

In many applied settings, it is important to have not just a point estimate of the wealth of a particular location, but also to have an understanding of the uncertainty associated with each point estimate. We are encouraged by the fact that we do not find evidence that the model performs any worse in poorer regions (Fig. S12), as occurs with nightlights data (*16*).

Disaggregating this error, we find that model error is lower when the target country is near to many countries with ground truth data used to train the model, and when there are many training observations nearby. This can be seen in Table S4, where we estimate the error of each individual 2.4km location $l$ by fitting a linear regression of the model's residual at $l$ (in the locations with ground truth data) on observable characteristics of $l$. We selected a broad set of observable characteristics that include: all of the features used in the predictive model (with the exception of the imagery-based features); how much "ground truth" training data was available near the spatial unit (such as the distance to the nearest DHS cluster); and country-level characteristics (such as average GDP per capita and continent dummy variables). We then regress the model error, in RWI units, of grid cell $l$ on the $l$'s vector of observable characteristics. We show the correlates of model error in Table S4, column 1.

[xviii] Data obtained from https://stats.oecd.org/Index.aspx?DataSetCode=PDB_LV. Of the 36 OECD member countries, 34 provide data on GDPpc. Of these, we exclude Luxembourg and Malta (which have only one and two geographic units, respectively). Ireland (6 units) and Lithuania (9 units) are also excluded since the spatial units listed in the GDPpc data do not match the spatial units listed in the corresponding OECD shapefiles. The remaining 30 countries contain 1690 administrative units, of which 1540 have GDPpc information.
[xix] The OECD's TL3 regions typically correspond to second-level administrative regions, with the exception of Australia, Canada and the United States. These TL3 regions are contained in a TL2 region, with the exception of the United States for which the Economic Areas cross the States' borders.

To better understand the sensitivity of these error estimates, we re-estimate the results in column 1 of Table S4 using different subsets of available predictors. Columns 2 and 3 of Table S4 indicate that while the point estimates $\beta$ depend somewhat on the other variables included in the regression, the qualitative patterns are the same. More importantly, we observe that the actual error estimates (for any given location *l*) are not very sensitive to the variables included in the model. For instance, Fig. S14 compares the error predicted by the model in column 1 of Table S4 (x-axis) against the error predicted by the two alternative specifications in columns 2-3. Fig. S14a shows the correlation between the median error of a country under the original specification and the median error of a country using a new specification that also includes the 100 satellite imagery features as predictors ($r$ = 0.770). Fig. S14b shows the correlation between the median country error under the original model and a model that only includes the set of features that were not used to estimate RWI ($r$ = 0.773).

More broadly, Fig. S6 and Fig. S13 indicate that models trained with data from a single country perform best when applied to countries with similar characteristics. To construct Fig. S13, we calculate the cosine similarity between all pairs of countries based on the country-level attributes listed in Table S4.[xx] We then show, for different thresholds of dissimilarity $d$, the average test error across all countries $c$ when the model is trained on countries at least $d$ dissimilar to $c$. For instance, when $d$ = 0.1, the model for each country $c$ is trained only on countries at least distance 0.1 from $c$.

Our objective in constructing the micro-estimates of model error is to provide policymakers and other users with a sense of where the model is accurate and where it is not. Fig. S3b provides a granular map of expected model error. We also provide country-level summary statistics of model error in Table S5 (i.e., the mean, median, and standard deviation of estimated model error in each country), to provide policymakers in specific countries with at-a-glance estimates of model performance.

## SM12. Absolute wealth estimates

The predictive models are trained to estimate the Relative Wealth Index (RWI) of each 2.4km grid cell. The RWI indicates the wealth of that location relative to other locations within the same country. However, certain practical applications require a measure of the *absolute* wealth of a region which can be more directly compared from one country to another.

To provide a rough estimate of the absolute per capita wealth of each grid cell, we use the technique proposed by Hruschka et al. (2015)(*42*) to convert a country's relative wealth distribution to a distribution of per-capita GDP. This method relies on three parameters to define the shape of the wealth distribution: the mean GDP per capita, as a measure of the central tendency ($GDP_c$); the Gini coefficient, as a measure of dispersion ($Gini_c$); and a combination of the Pareto and log-normal distributions that are used to estimate skewness. Specifically, our Absolute Wealth Estimate (AWE) of a grid cell $i$ in country $c$ is defined by:

[xx] Specifically, the features are: area, population, island, landlocked, distance to the closest country with DHS, number of neighboring countries with DHS, GDP per capita, and Gini coefficient.

$$AWE_{ic} = rank_{ic} * \frac{GDP_c}{\frac{1}{n}\sum_j ICDF_c(rank_j)}$$

where $rank_{ic}$ is the rank of each grid cell's RWI (relative to other cells in $c$), $GDP_c$ is the mean wealth per capita of $c$, and $ICDF_c$ is the inverse cumulative distribution of wealth, which is parameterized exactly following Hruschka et al. (2015).[xxi] We collect indicators of each country's Gini coefficient and mean per capita GDP from the sources listed in Table S6, and use it to produce the Absolute Wealth Estimates (AWE) shown in Fig. a.

This conversion requires strong parametric assumptions about the national distribution of wealth based on information about the average wealth and wealth inequality in each country. These assumptions are not justified in many countries, particularly where Gini estimates are unreliable, or when the ICDF approximation is a poor fit. Thus, the AWE estimates should be treated with more caution than the RWI estimates, which were carefully validated with several different sources of independent survey data.

Fig. S15 shows the global distribution of (predicted) absolute wealth, as derived from the Relative Wealth Index using the above procedure. The figure compares the predicted wealth distribution based on our method to the global income distribution in 2013, as independently estimated by Hellebrandt and Mauro (2015)(*43*) using household income surveys for more than a hundred countries that were collected through the Luxembourg Income Study. As expected, the average wealth distribution, which is a measure of per capita GDP, is uniformly higher than the estimated income distribution, which reflects actual family incomes (i.e., total economic output does not translate directly to better family outcomes).

## SM13. Targeting simulations

To illustrate one practical use case for these micro-estimates, we simulate the scenario in which an anti-poverty program administrator has a fixed budget to distribute to a country's population. Following Ravallion (*25*) and Elbers et al. (*2*), we assume that the program will be geographically targeted, such that all individuals within targeted regions will receive the same transfer. Our analysis compares the performance of several different approaches to geographic targeting in Togo (Table S7) and Nigeria (Table S8), with a subset of these results summarized in Table 1. Performance is evaluated using recent nationally-representative household survey data collected in each country (see above for a description of the EHCVM and NLSS datasets used to evaluate targeting outcomes).

In both Table S7 (for Togo) and Table S8 (for Nigeria), Panel A simulates geographic targeting using the high-resolution ML estimates. The first row simulates a scenario in which cash is transferred to households located in the poorest 2.4km tiles of the country; the second row simulates distribution to the households located in the poorest administrative units of the country (the canton is the smallest administrative unit in Togo and the ward is the smallest administrative

[xxi] For the Pareto distribution, $ICDF_c$ is the inverse cumulative distribution function with shape parameter $\alpha = (1 + Gini_c)/(2\ Gini_c)$, using a threshold of $\left[1 - \left(\frac{1}{\alpha}\right)\right]$. Otherwise, $ICDF_c$ is for a log-normal distribution based on a normal distribution with a mean of: $Ln(GDPpc_c) - \sigma^2/2$, where $\sigma = \sqrt{2} * probit(\frac{Gini_c+1}{2})$.

unit in Nigeria), where the wealth of the administrative unit is calculated as the population-weighted average of the RWI of all tiles in that unit. The first column indicates the number of unique tiles in each country; the second and third columns simply indicate that every spatial unit (tile or canton/ward) has a corresponding wealth estimate. Column 4 indicates the number of spatial units for which ground truth data exist (in the EHCVM or NLSS), and column 5 counts the number of spatial units for which both ML estimates and ground truth data exist. Column 6 indicates the number of households that exist in those spatial units for which there are both ML estimates and ground truth data. This set of households is then used to measure the correlation between the ground truth wealth of each household (i.e., "true wealth") and the ML estimate of the wealth of the spatial unit in which that household is located (i.e., "predicted wealth"), which is reported in Column 7.[xxii] In subsequent columns, we assume that the government has a fixed budget which allows it to only target 25% or 50% of the population. We consider the "true poor" to be the 25% or 50% of households in the ground truth survey with the lowest household asset index. In Panel A, the targeting mechanism we simulate selects the 25% or 50% "predicted poor" households, where the prediction is based on the ML estimate of wealth assigned to the spatial unit in which each household is located. In instances where including one additional spatial unit would imply that more than 25% or 50% of households would receive benefits, households from that region are randomly selected to ensure that exactly 25% or 50% of households receive benefits. Columns 8 and 9 report the accuracy of this targeting mechanism; columns 10 and 11 report the precision and recall.[xxiii]

For comparison, Panels B-D simulate alternative geographic targeting approaches that a policymaker might rely on in the absence of comprehensive household-level data on poverty status, as is the case in many LMICs (*44*). In these simulations, we assume that the policymaker does not have access to the ML micro-estimates of RWI or the ground truth data from the EHCVM/NLSS that is used to evaluate their allocation decisions. Instead, the policymaker designs a geographic targeting policy based on the most recent DHS survey, which was conducted in 2018 in Nigeria and 2013-14 in Togo.

In Panel B, each row corresponds to targeting at a different level of geographic aggregation. For instance, the row labeled "prefecture average" in Panel B of Table S7 assumes that the program will be targeted at the prefecture level, the 2nd-level administrative region in Togo, such that either all households in the prefecture will receive benefits or none will. Subsequent rows allow for targeting at smaller geographic units. The columns in Panel B are organized similarly to Panel A. Note, however, that it is no longer the case that each spatial unit will necessarily have a "predicted wealth" value. For instance, in the Canton targeting row of Panel B (Column 2) indicates that only 185 cantons have one or more surveyed households in the most recent DHS (i.e., only 47.8% of all cantons). Columns 4-6 are analogous to Panel A. In Column 7, the "predicted wealth" of each household is the average wealth of all households in that region from the most recent DHS. In subsequent columns, the targeting mechanism selects the 25% or 50%

[xxii] This table reports the correlation in wealth at the *household* level, with one observation per household, using the household survey weights in the EHCVM/NLSS. This approach is most consistent with the targeting simulations, which require that the policymaker estimate each household's wealth. This approach is different than that taken to construct Fig. 4, which shows correlations at the *tile* level, with one observation per tile, which is consistent with the earlier objective of evaluating the accuracy of the ML estimates at the geographic level.
[xxiii] Precision and Recall are always equal in these targeting simulations because the fixed budget constraint implies that each additional targeting error creates exactly one new false positive and one new false negative.

"predicted poor" households, where the prediction is based on the average wealth of all households in that region from the most recent DHS.

Panel C simulates targeting in a similar manner to Panel B, with one important difference: In cases where a geographic unit has no surveyed households in the most recent DHS (e.g., 52.2% of all prefectures in Togo), we impute the wealth of that geographic unit by taking the average DHS RWI of all households in the geographic unit closest to the household *i*. The imputation on Panel C addresses a fundamental limitation of Panel B, which would otherwise leave policymakers without a mechanism to determine budget allocation in large regions of the country where survey data do not exist.

Panel D simulates a "nearest neighbor" approach to targeting, where the wealth of a household *i* is inferred based on the average wealth of the households in the DHS cluster physically closest to *i*, irrespective of whether those nearest neighbors are located in the same administrative unit as *i*.

The targeting simulations highlight three main results. First, the ML estimates allow for geographic targeting at a level of spatial resolution that would not be possible with traditional survey-based data. As highlighted in prior work (*25*, *26*, *2*), geographic disaggregation can produce substantial welfare gains. The gains to disaggregation are quantified in the last several columns of Table S7 and Table S8, which highlight how targeting at the tile level increases both precision and recall – i.e., it reduces both errors of exclusion and errors of inclusion – relative to the other targeting options that provide 100% coverage.[xxiv] In practice, it may be logistically challenging to deliver benefits to such small geographic units, but recent and ongoing work that uses mobile money to deliver cash transfers directly to beneficiaries suggests that this type of approach may soon become feasible (*1*).

Second, even if the delivery of benefits will be based on administrative divisions, we find that admin-region targeting based on the ML estimates performs at least as well as – and often better than – admin-region targeting based on recent nationally representative household surveys (i.e., the comparison of the last row of Panel A to the last row of Panel C or Panel D). This is because the ML estimates can be used to construct accurate estimates of the wealth of 100% of administrative units. By contrast, the DHS only surveyed households in 185 (47.8%) cantons in Togo, and only 1218 (13.8%) wards in Nigeria. Thus, a geographic targeting approach relying on the DHS data alone would either require implementation at a larger administrative unit, or would require some other form of imputation into unsurveyed regions (as is the case in Panels C and D) – both of which adjustments reduce the effectiveness of geographic targeting.

Third, and echoing previous results, the ML estimates are accurate at estimating household wealth (column 7 of Table S7 and Table S8), and are at least as accurate as household wealth estimation based on recent DHS data. In this sense, Table S7 and Table S8 provide a conservative estimate of the gains from using the ML estimates for geographic targeting. Many LMICs do not have a recent nationally representative household survey available; for instance,

[xxiv] In Panel B of Table a, the "canton targeting" approach slightly outperforms the tile-level targeting, but as we discuss below, the approach described in Panel B could not be used to target the majority of cantons in Togo, since only 47.8% of cantons contain households that participated in the DHS.

only 24 of 135 LMICs have conducted a DHS since 2015. For such countries, these micro-estimates create options for geographic targeting that might otherwise not exist.

Finally, we note that the above discussion compares universal geographic targeting using the ML estimates to universal geographic targeting using recent DHS data, such that all individuals in a targeted region receive uniform benefits. In practice, most real-world programs are more nuanced, and rely on additional targeting criteria (such as proxy means tests and participatory wealth rankings) to determine program eligibility. These additional criteria would be expected to increase the performance of all methods listed in Table S7 and Table S8; we do not simulate those changes to better highlight the gains from geographic disaggregation.

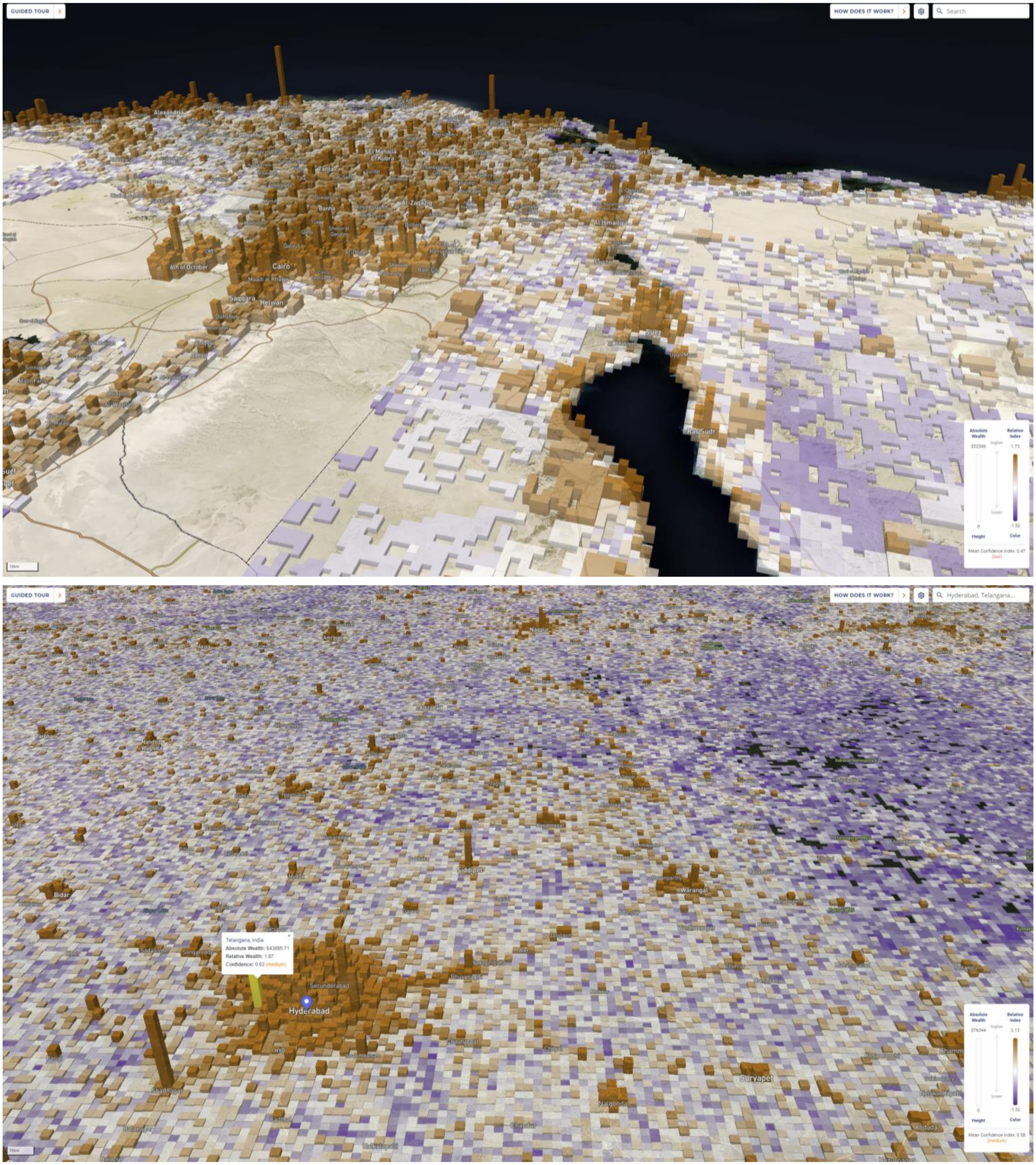

**Fig. S1 | Screenshots of the interactive data visualization**. Each grid cell corresponds to a 2.4km grid cell. Absolute wealth (in dollars) indicated by the height of the grid cell. Relative wealth (relative to other cells in that country) indicated by colors ranging from blue (poorest) to red (wealthiest). **a)** Region around the Suez canal**. b)** Region around Hyderabad, India.

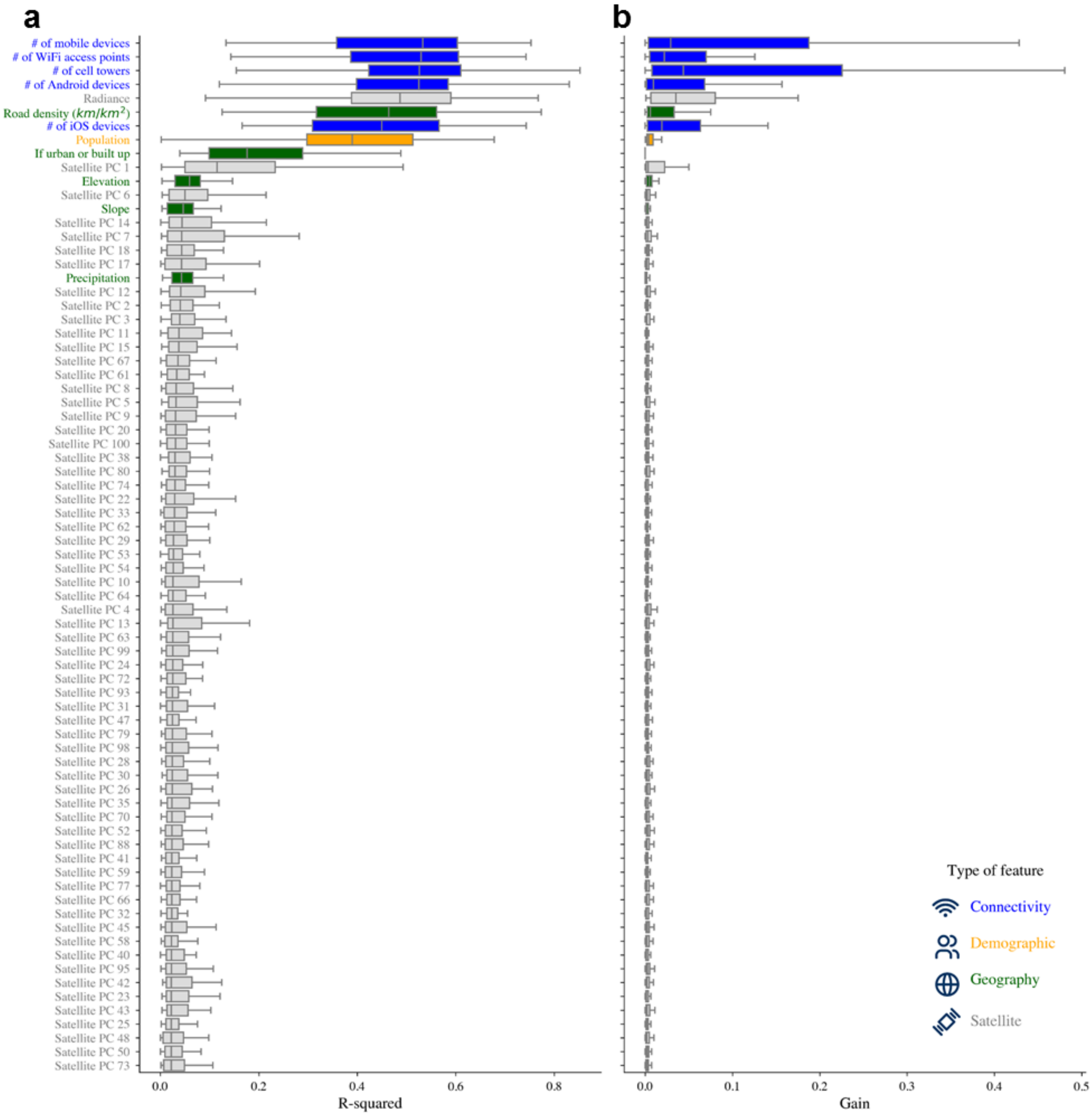

**Fig. S2 | Which input data are most useful?** Two different measures of the importance of each input variable in predicting sub-regional wealth. We show the distribution of feature importances for each feature as a boxplot, where the distribution is obtained from training 56 country-specific models with 5-fold cross-validation. **a)** The $R^2$ value from a univariate regression of wealth on each feature. **b)** Gain measures the total contribution of each feature to the final fitted model. Details on each variable are provided in Table S2. Box plots indicate median (center line), interquartile range (shaded box), and 1.5x interquartile range (whiskers).

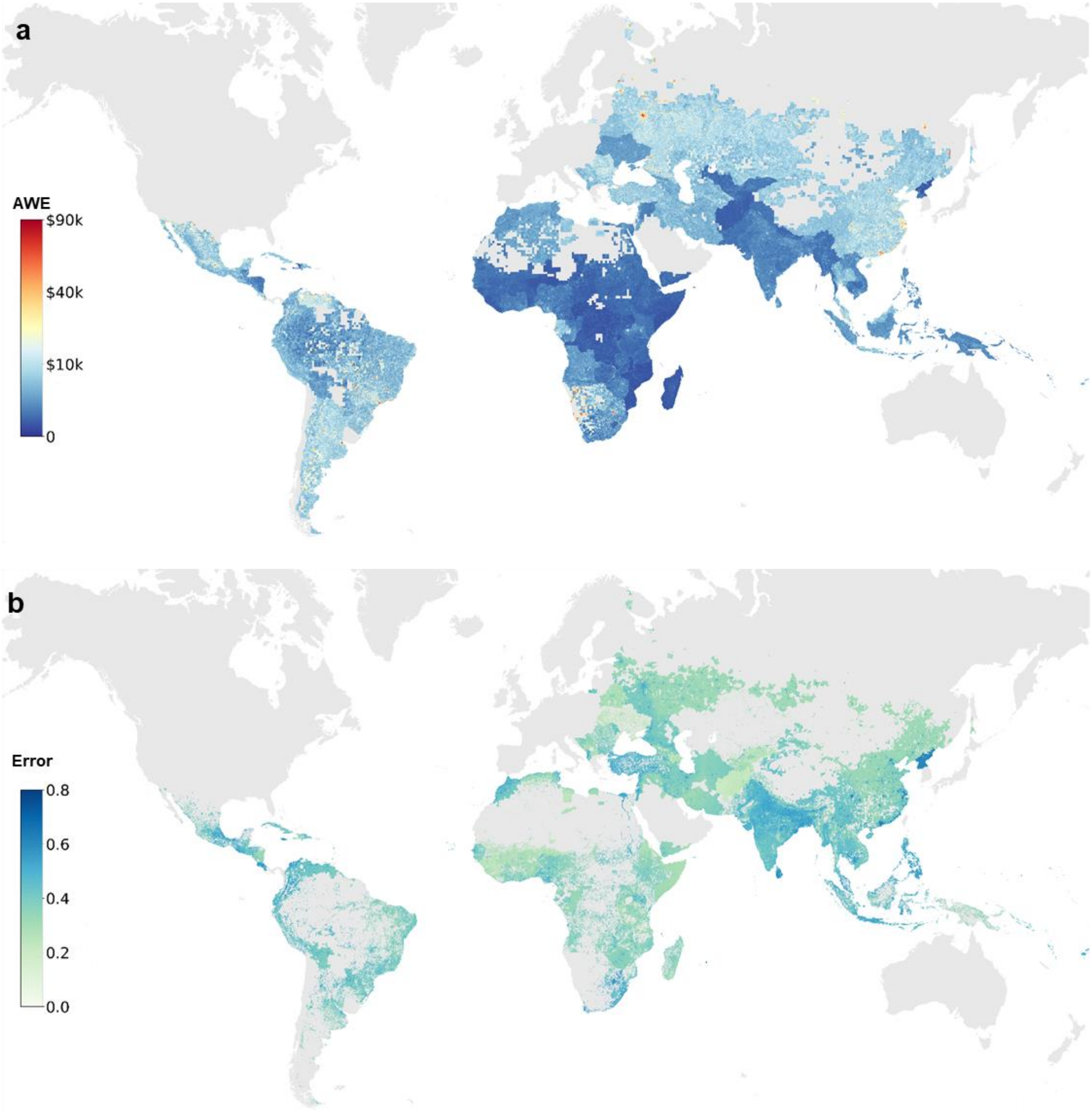

**Fig. S3 | Estimates of absolute wealth and of model error. a)** Absolute wealth estimates (AWE), measured as the average GDP per capita of households in each grid cell. RWI estimates are converted to AWE estimates using information about the income distribution of each country. **b)** Predicted absolute error of each grid cell, based on a regression of absolute model residual on observable characteristics of each grid cell.

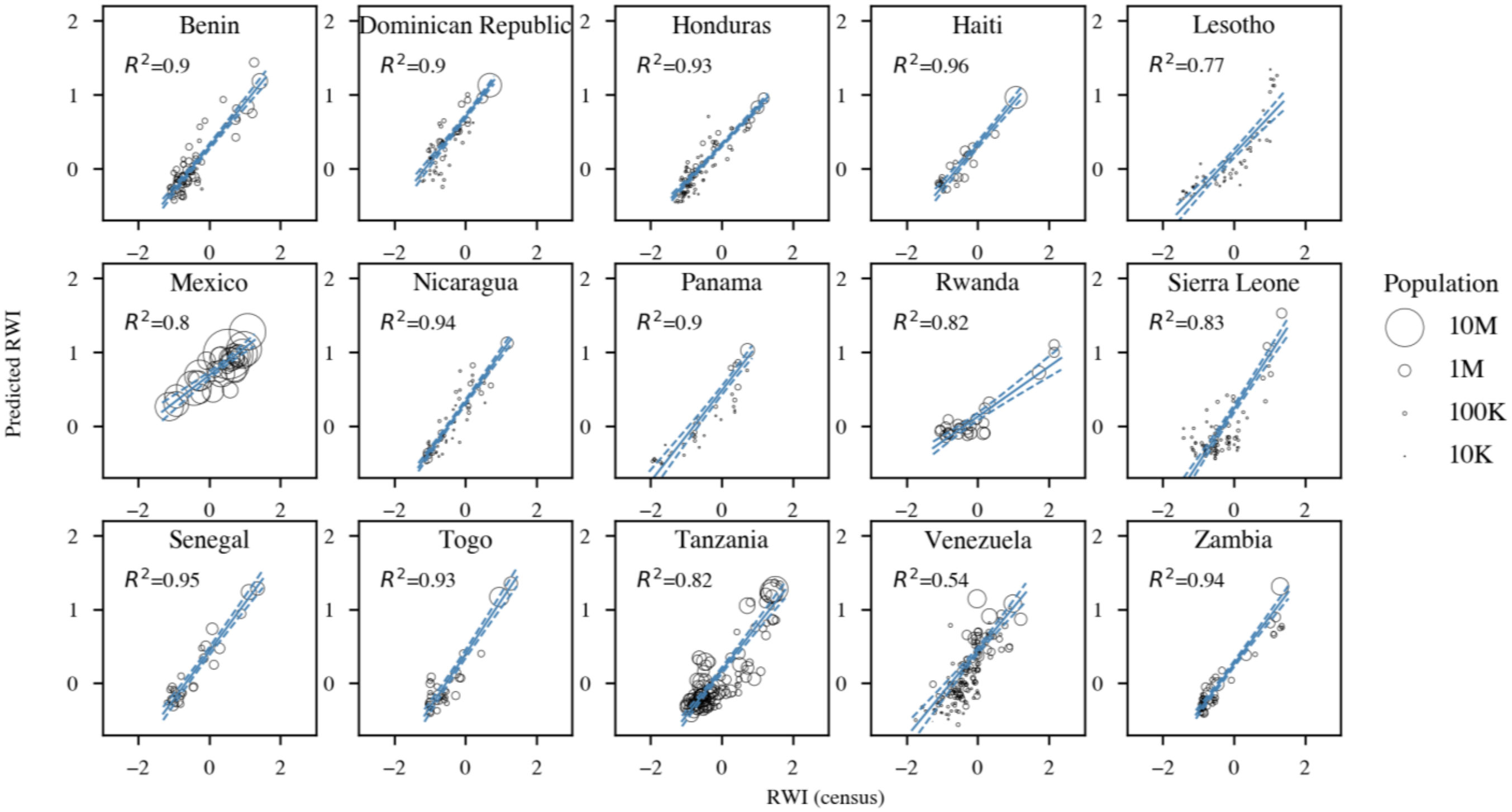

**Fig. S4 | Model validation using census data in low- and middle-income countries.** For each of 15 countries with publicly available census data, we compare the average RWI of each second-level administrative region, as predicted by the ML model, to the average wealth captured in the census (see Methods). Each dot represents an administrative region, sized by population. Blue line indicates population-weighted regression line, with 95% confidence intervals as dashes. Average $R^2$ across all models is 0.86.

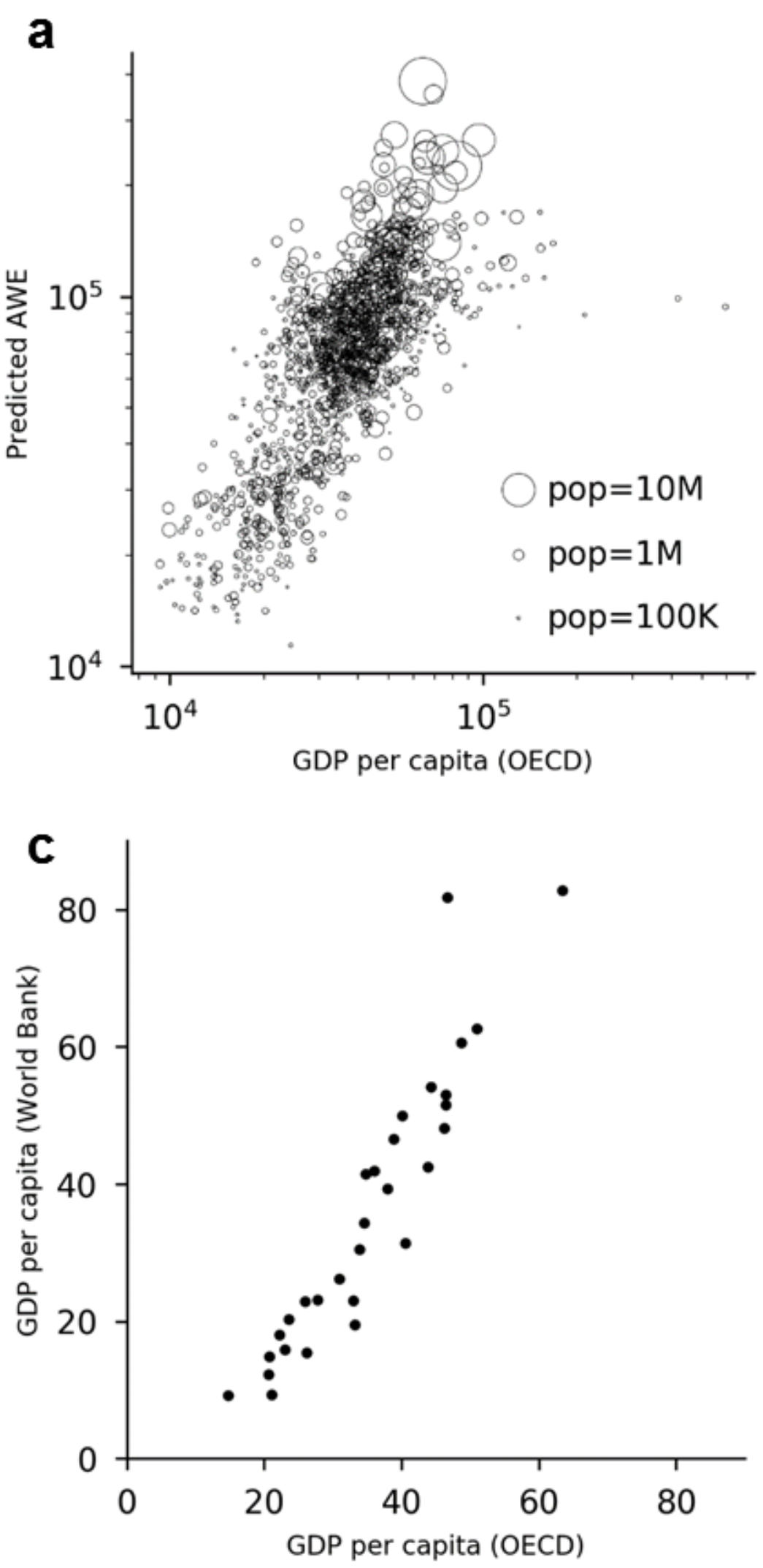

**b**

| | *Country* | *Year* | *# Units* | *$R^2$ (AWE)* | *$R^2$ (RWI)* |
|---|---|---|---|---|---|
| *1* | Austria | 2017 | 35 | 0.488 | 0.536 |
| *2* | Belgium | 2018 | 44 | 0.621 | 0.735 |
| *3* | Bulgaria | 2017 | 28 | 0.779 | 0.648 |
| *4* | Croatia | 2017 | 21 | 0.502 | 0.663 |
| *5* | Czechia | 2017 | 14 | 0.914 | 0.818 |
| *6* | Denmark | 2018 | 11 | 0.528 | 0.652 |
| *7* | Estonia | 2018 | 5 | 0.996 | 0.977 |
| *8* | Finland | 2017 | 19 | 0.896 | 0.887 |
| *9* | France | 2016 | 101 | 0.070 | 0.402 |
| *10* | Germany | 2017 | 401 | 0.297 | 0.308 |
| *11* | Greece | 2017 | 52 | 0.046 | 0.305 |
| *12* | Hungary | 2018 | 19 | 0.201 | 0.296 |
| *13* | Italy | 2017 | 110 | 0.000 | 0.022 |
| *14* | Japan | 2016 | 47 | 0.259 | 0.282 |
| *15* | Rep. of Korea | 2018 | 17 | 0.021 | 0.032 |
| *16* | Latvia | 2017 | 6 | 0.951 | 0.981 |
| *17* | Netherlands | 2018 | 40 | 0.107 | 0.248 |
| *18* | New Zealand | 2017 | 12 | 0.495 | 0.570 |
| *19* | Norway | 2018 | 18 | 0.456 | 0.424 |
| *20* | Poland | 2017 | 73 | 0.609 | 0.634 |
| *21* | Portugal | 2017 | 25 | 0.787 | 0.709 |
| *22* | Romania | 2017 | 42 | 0.899 | 0.820 |
| *23* | Slovakia | 2018 | 8 | 0.881 | 0.904 |
| *24* | Slovenia | 2018 | 12 | 0.744 | 0.629 |
| *25* | Spain | 2017 | 59 | 0.441 | 0.457 |
| *26* | Sweden | 2017 | 21 | 0.702 | 0.607 |
| *27* | Switzerland | 2017 | 26 | 0.289 | 0.358 |
| *28* | Turkey | 2018 | 81 | 0.744 | 0.773 |
| *29* | United Kingdom | 2018 | 179 | 0.034 | 0.064 |
| *30* | USA | 2018 | 179 | 0.259 | 0.298 |
| | ***Mean (30 OECD countries)*** | | | ***0.500*** | ***0.535*** |

**Fig. S5 | Model validation in high-income nations.** Figures compare the model's estimates of wealth to data provided by the OECD for 30 member countries. **a)** The 30 nations contain 1540 unique second-level administrative regions, each of which is represented by a dot that is sized proportional to the population of the region. Figure shows the OECD's estimate of per capita GDP for the region (x-axis) vs. the Absolute Wealth Estimates (AWE) of the region generated by the ML model. Population-weighted regression line $R^2 = 0.59$. **b)** We separately calculate, for each of the 30 OECD countries with available GDP data, the $R^2$ that results from regressing predicted AWE on GDPpc, across all admin-2 regions within each country. **c)** The estimate of a country's GDPpc from the World Bank, which forms the basis for the AWE estimates, is generally larger than the average regional GDPpc as reported in the OECD data. Values on axes represent thousands of US Dollars.

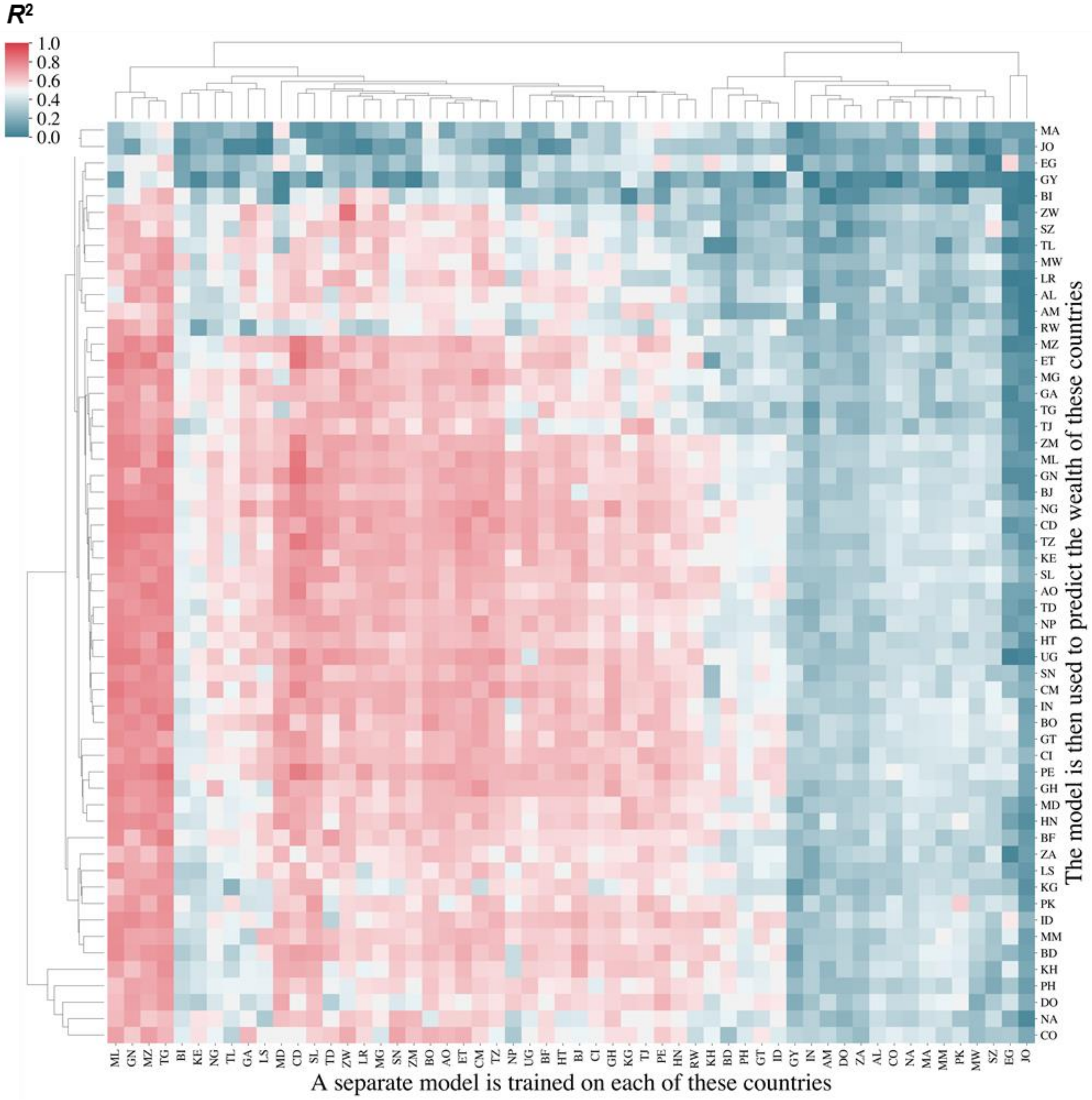

**Fig. S6 | Geographic generalizability of wealth predictions**. For each of the 56 countries with ground truth wealth data, a separate model is trained using data from just that country (the columns in the above matrix). Those models are then tested on previously unseen data from each of the countries (the rows in the matrix). Colors indicate the $R^2$ between the model's predictions and ground truth. Models generally perform better on nearby and similar countries. Rows and columns are ordered using a hierarchical clustering algorithm (UPGMA).

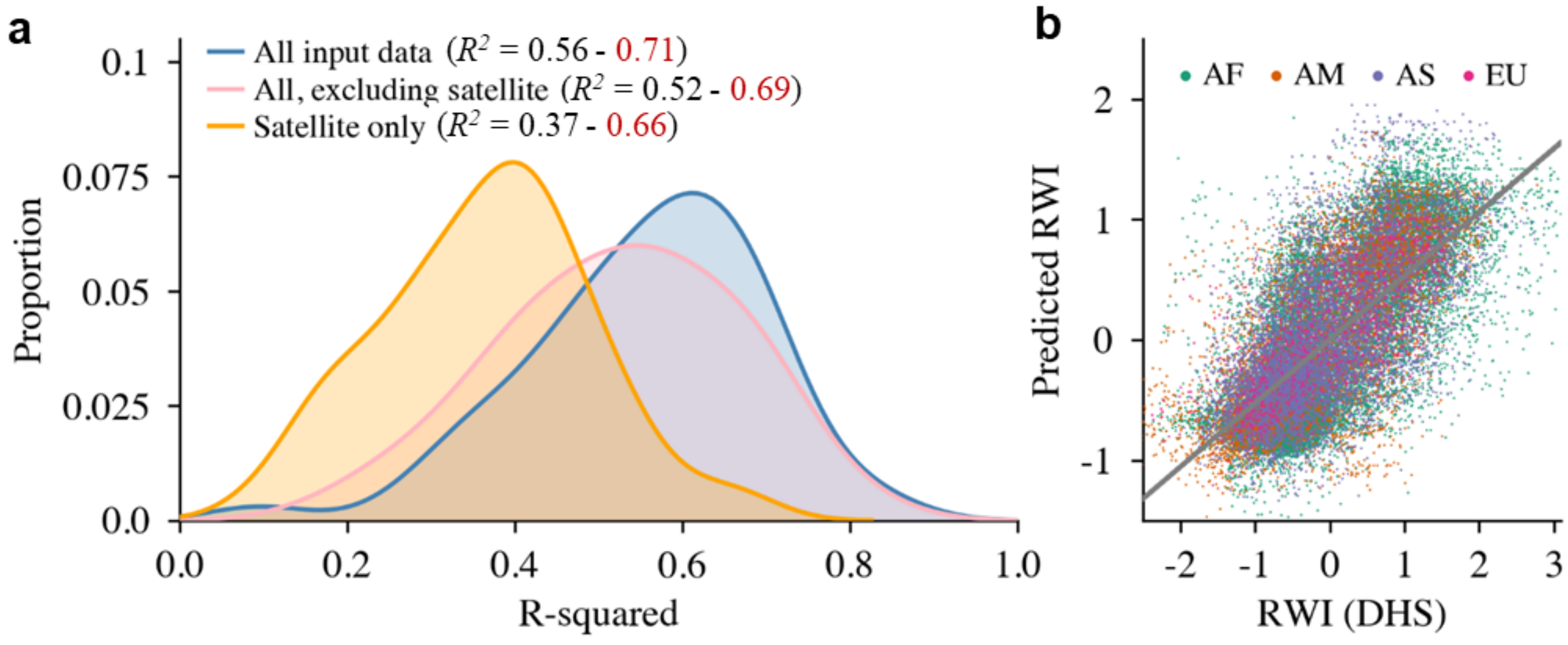

**Fig. S7 | Models trained only on satellite data do not perform as well as models that include other input data. a)** The distribution of performance across the 56 LMIC's, measured using spatially-stratified cross-validation, is shown as three kernel density plots, one for each subset of input data. The legend reports the average performance ($R^2$) in black, and the average performance using standard cross-validation in red (to facilitate comparison to prior work). **b)** Scatter-plot shows relationship between the actual wealth index (from survey data) and the predicted wealth index (output by the model), using all 66,819 labeled survey locations on four continents (AF=Africa, AM=Americas, AS=Asia, EU=Europe).

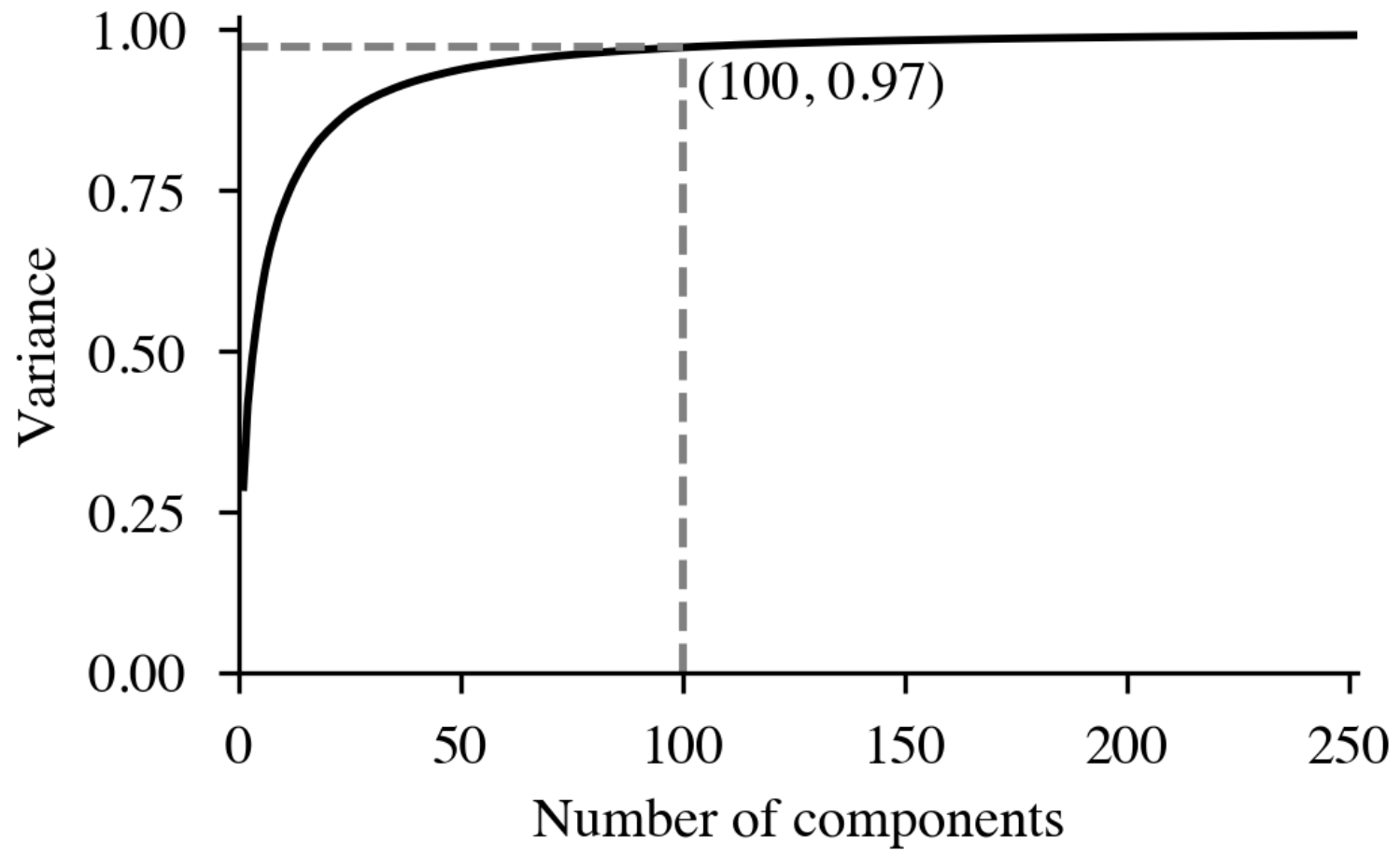

**Fig. S8 | Feature engineering from satellite imagery.** To reduce the dimensionality of the raw satellite imagery, we first use a neural network to extract 2048 features from the data (see Methods), and then apply principal component analysis (PCA) to the set of 2048 features. Figure shows the cumulative proportion of variance explained by the first *k* principal components. Our final model uses 100 components, which cumulatively explain 97% of the total variance of the 2048-dimensional image features.

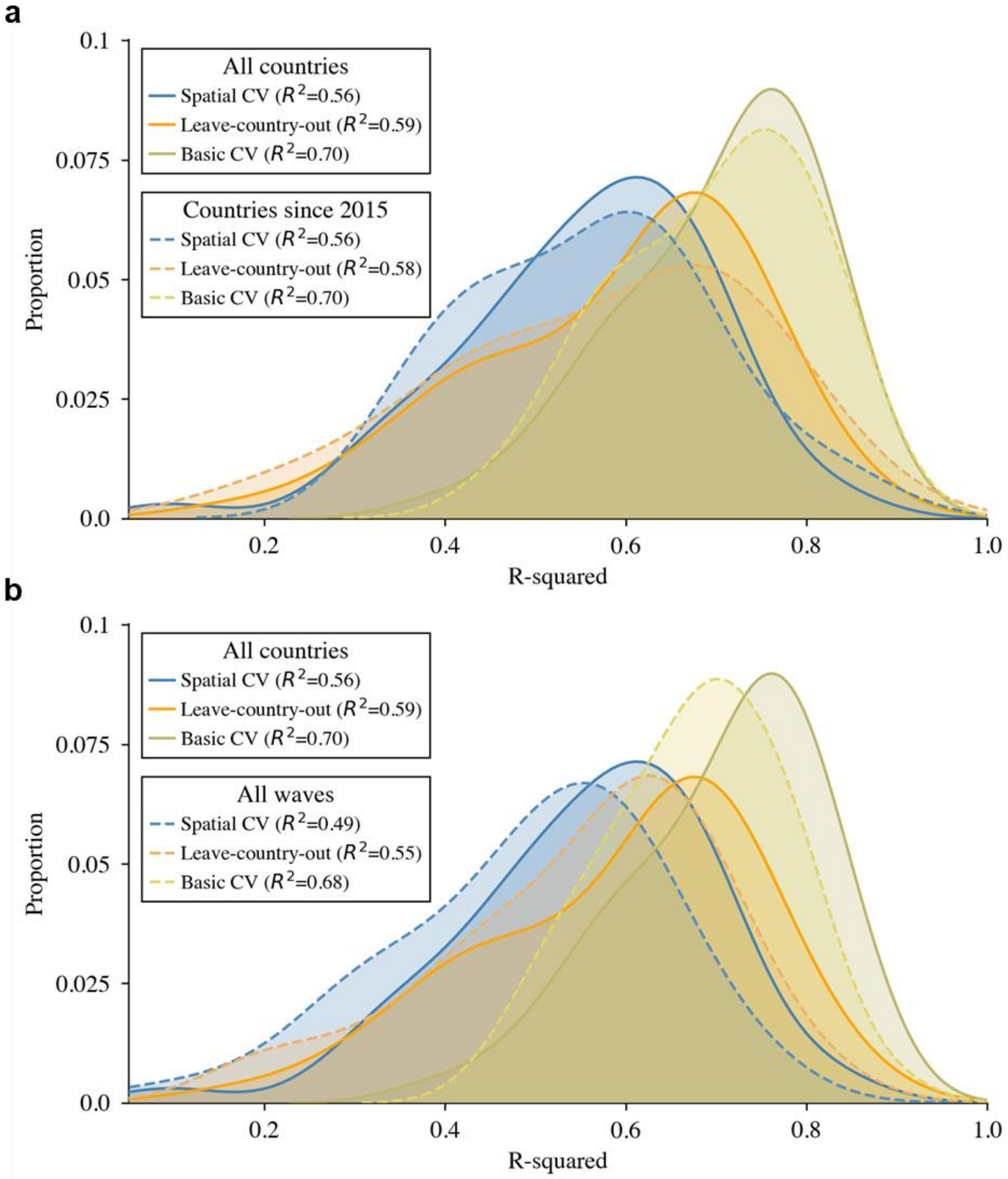

**Fig. S9 | Model performance when trained using surveys from different periods in time. a)** Performance using recent data only. The solid lines (labeled "All countries") reproduce the analysis of Fig. 3a, and show the distribution of model performance for a model trained on 56 countries with DHS data, using 3 different approaches to model cross-validation. The dashed lines indicate the performance for a model that is trained on the subset of 24 countries where DHS data was collected in 2015 or later. **b)** Performance using all available survey waves. Several countries have conducted multiple DHS surveys since 2000. The figure compares the main model's performance (using 56 survey-waves from 56 countries) to the performance of a model trained and evaluated on all available DHS data (117 survey-waves from 56 countries).

a

| Country | Code | DHS Year | Census | $R^2$ |
|---|---|---|---|---|
| **Benin** | BJ | 2017-18 | 2013 | 0.90 |
| **Dom. Rep.** | DO | 2013 | 2010 | 0.90 |
| **Haiti** | HT | 2016-17 | 2003 | 0.96 |
| **Honduras** | HN | 2011-12 | 2001 | 0.93 |
| **Lesotho** | LS | 2014 | 2006 | 0.77 |
| **Rwanda** | RW | 2014-15 | 2012 | 0.82 |
| **Senegal** | SN | 2016 | 2002 | 0.95 |
| **Sierra Leone** | SL | 2013 | 2004 | 0.83 |
| **Tanzania** | TZ | 2015-16 | 2012 | 0.82 |
| **Togo** | TG | 2013-14 | 2010 | 0.93 |
| **Zambia** | ZM | 2018 | 2010 | 0.94 |

b

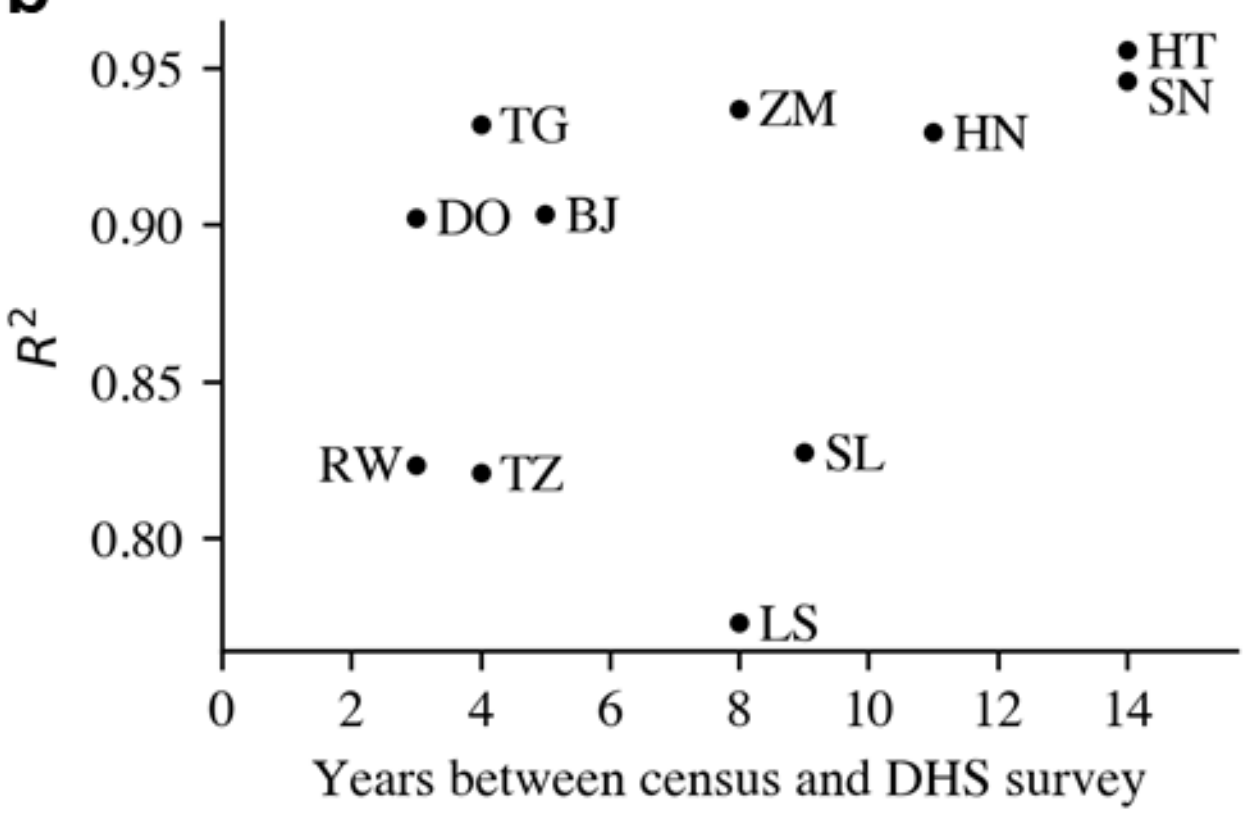

**Fig. S10 | Validation accuracy for 11 countries with DHS and census data. a)** Of the 56 countries used to train the model, 11 have publicly available census data with asset information. The table indicates the dates of the most recent DHS survey and census in each country, as well as the correspondence ($R^2$) between the model predictions and the census data (see also Fig. **S4**). **b)** Figure illustrates that there is no clear relationship between the gap in years between the most recent DHS survey and census (x-axis) against the validation accuracy of the model, for each of these 11 countries.

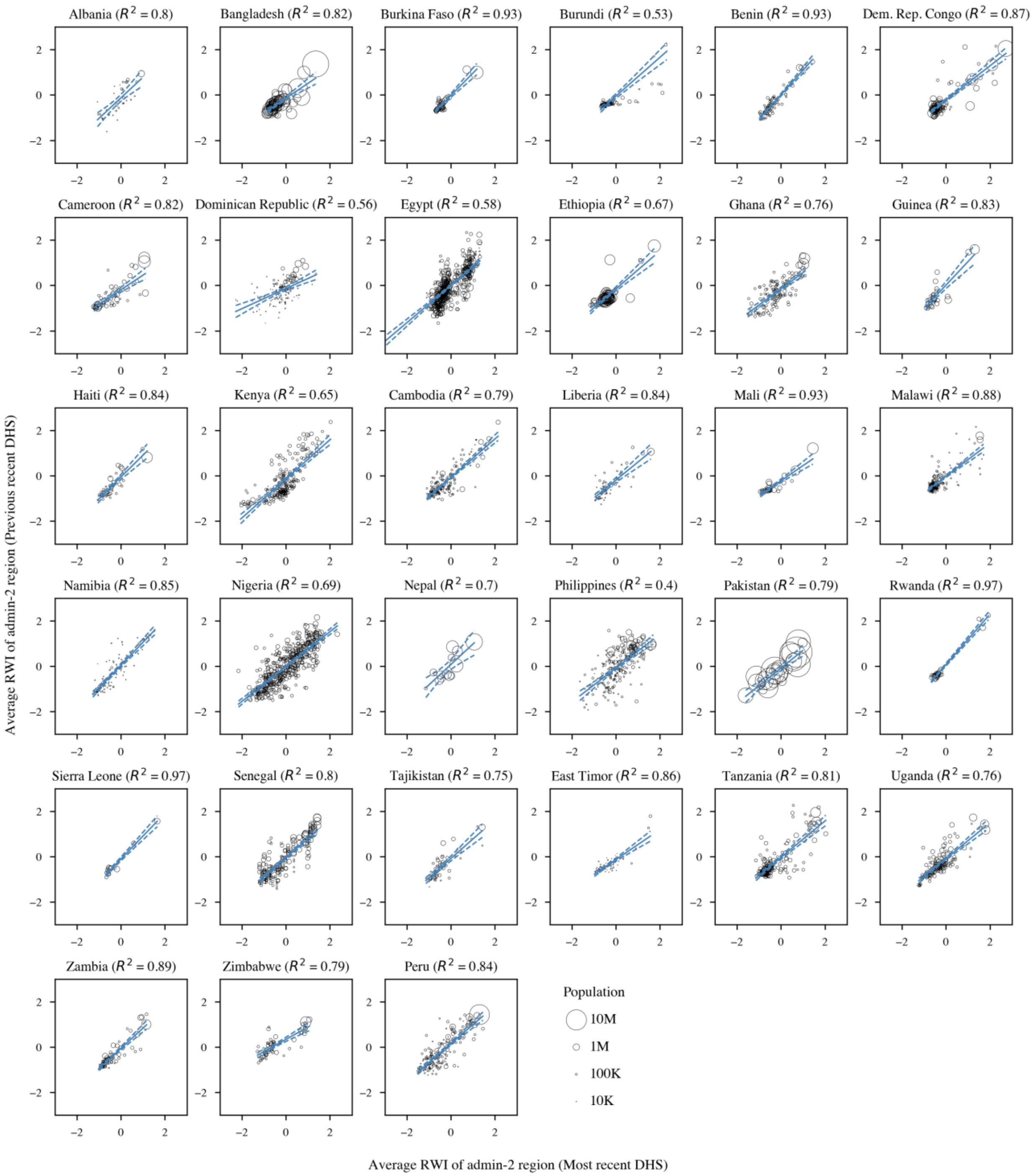

**Fig. S11 | Temporal stability of within-country wealth over time.** For each country with two or more DHS since 2000, each subfigure plots the relationship between the average RWI of each 2nd-level administrative unit as computed from the most recent DHS (x-axis) against the average RWI of the same unit as computed from the previous DHS. Each circle represents an administrative region. Blue line indicates population-weighted regression line, with 95% confidence intervals as dashes. Median (mean) $R^2$ across all countries is 0.81 (0.78).

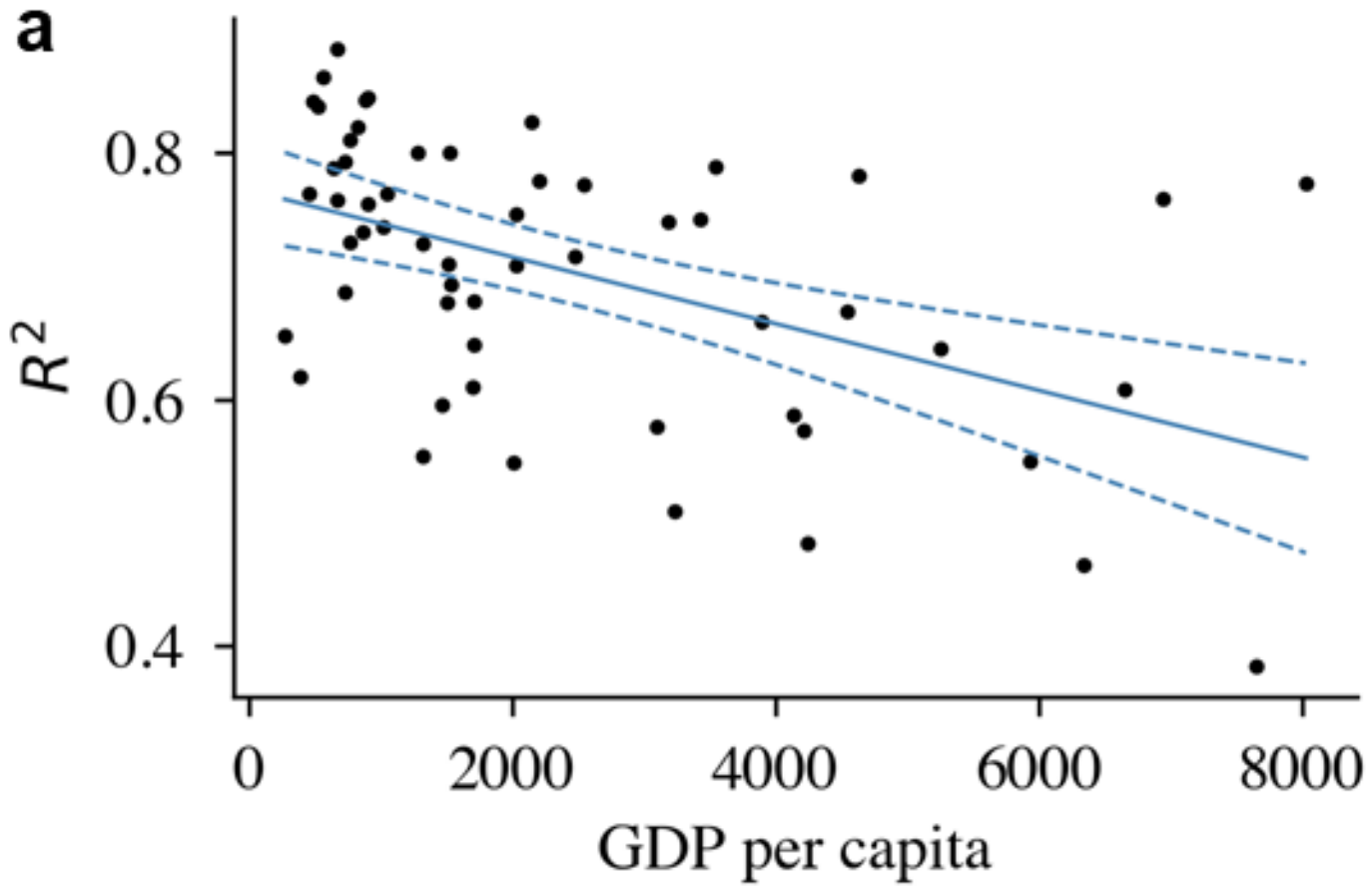

b

| | $\beta$ | SE |
|---|---|---|
| Low income | -0.0228 | 0.041 |
| Upper middle income | -0.2084*** | 0.052 |
| Gini | -0.3560 | 0.283 |
| ln(Country population) | -0.0021 | 0.013 |
| Is island | -0.0336 | 0.057 |
| Is landlocked | 0.0083 | 0.040 |
| Is Africa | 0.0490 | 0.106 |
| Is Asia | -0.0633 | 0.097 |
| Is America | 0.1085 | 0.109 |
| Constant | 0.7570*** | 0.230 |
| Observations | 56 | |
| $R^2$ | 0.404 | |

*Note:* Lower-middle income is omitted country category.
*significant at 10 percent; ** 5 percent; *** 1 percent.

**Fig. S12 | Model performance and country characteristics.** a) For each of the 56 countries with ground truth data from the DHS, the figure plots the country-level $R^2$ (measured using basic 5-fold cross-validation) against that country's GDP per capita, as measured in Table S6. **b)** Coefficients and standard errors from a regression of the country-level $R^2$ on country-level characteristics, for the 56 countries with ground truth data, indicates that model performance is slightly worse in upper middle-income countries (relative to the omitted category of lower-middle income countries, but is not significantly different in low-income countries or specific continents.

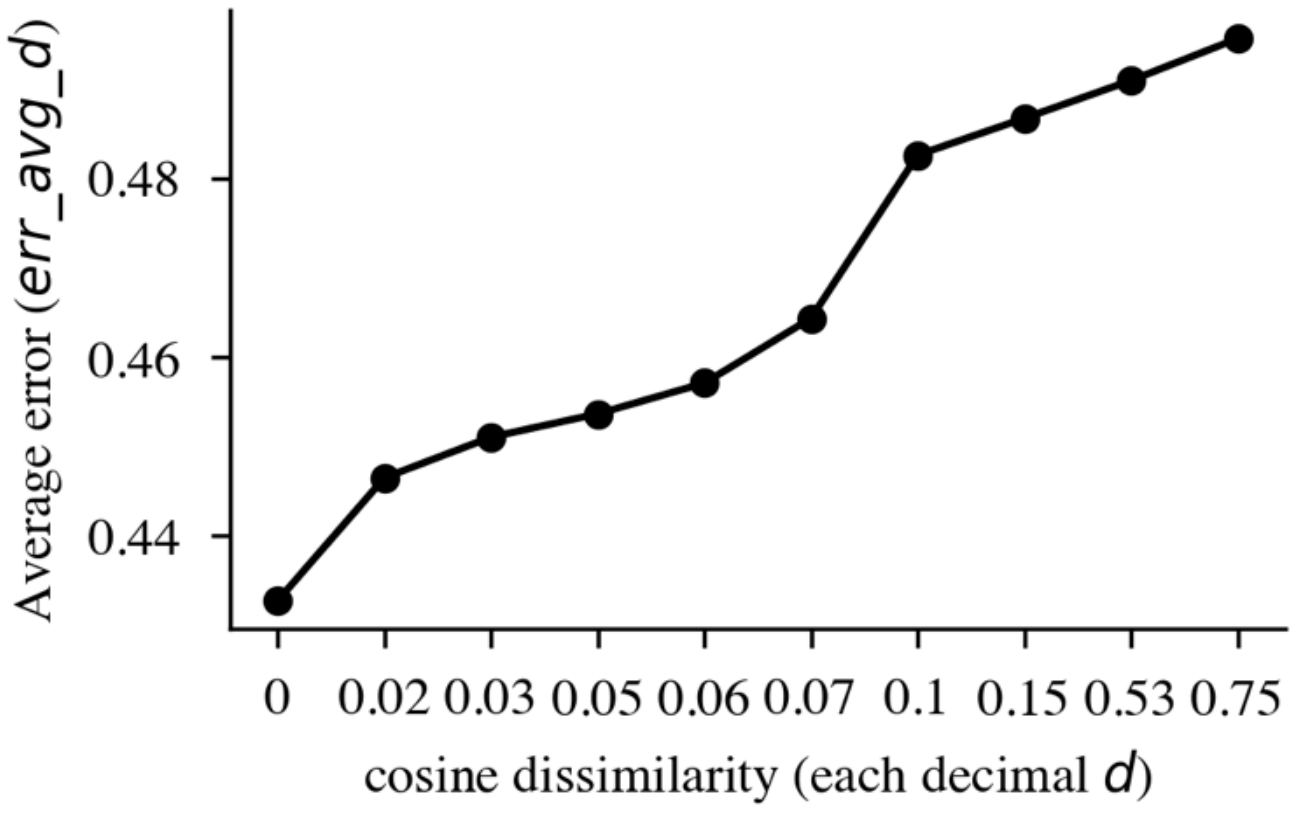

**Fig. S13 | Models perform better on countries with similar characteristics.** X-axis shows the 10 deciles of the cosine dissimilarity distribution (i.e., one minus cosine similarity). Y-axis indicates the average prediction error across test countries, where a separate model for each test country is trained using data from countries at least *d* dissimilar to the test country.

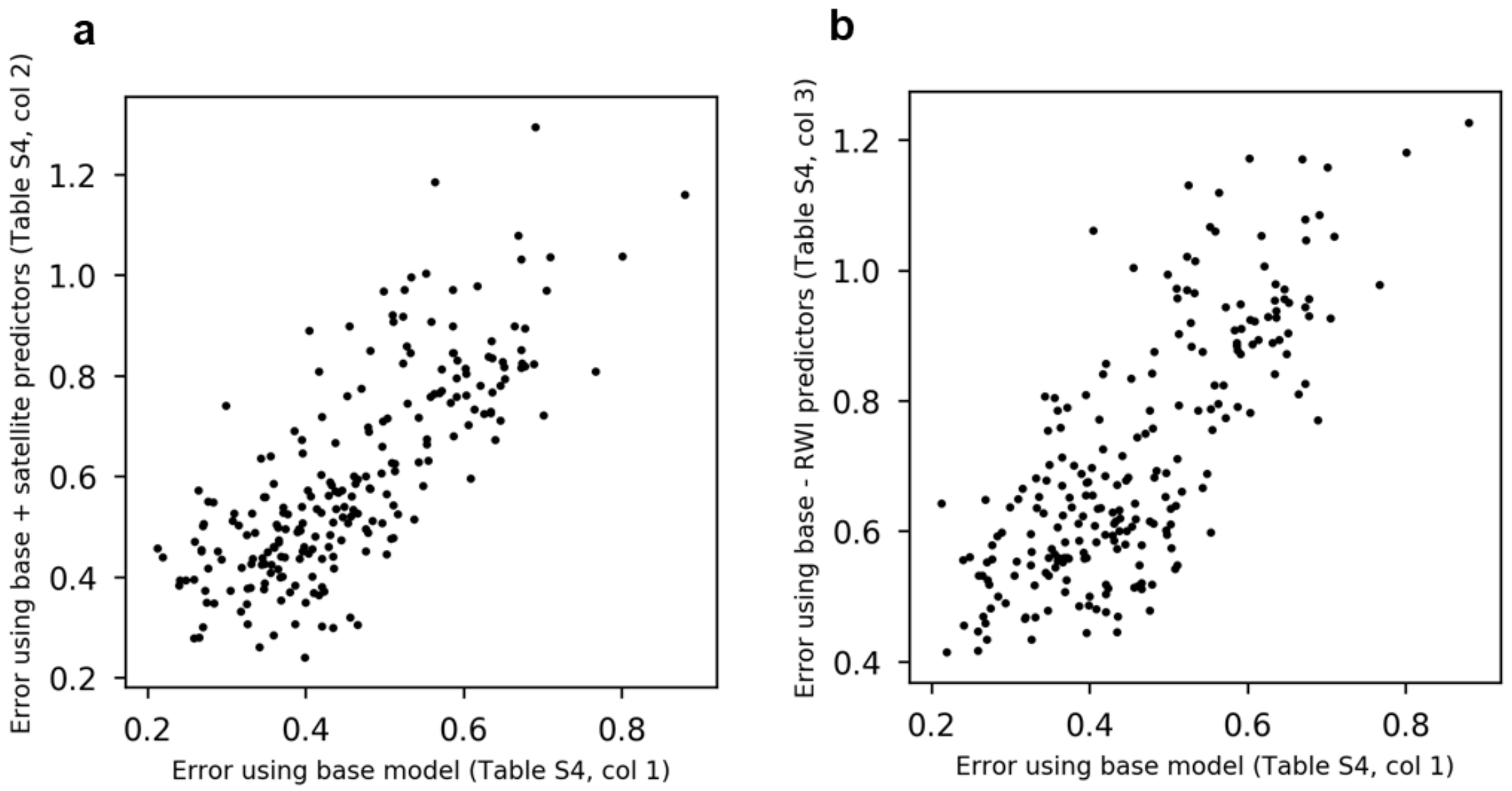

**Fig. S14 | Stability of error estimation to different model specifications.** Figure compares the median error of all grid cells in a country from the base model (defined by column 1 of Table S4, and plotted on the x-axis of both figures) to two alternative model specifications (plotted on the y-axis). Each dot represents a country. **a)** Alternate model includes 100 satellite-based features (defined by column 2 of Table S4). **b)** Alternate model limited to only features not used to predict RWI (defined by column 3 of Table S4).

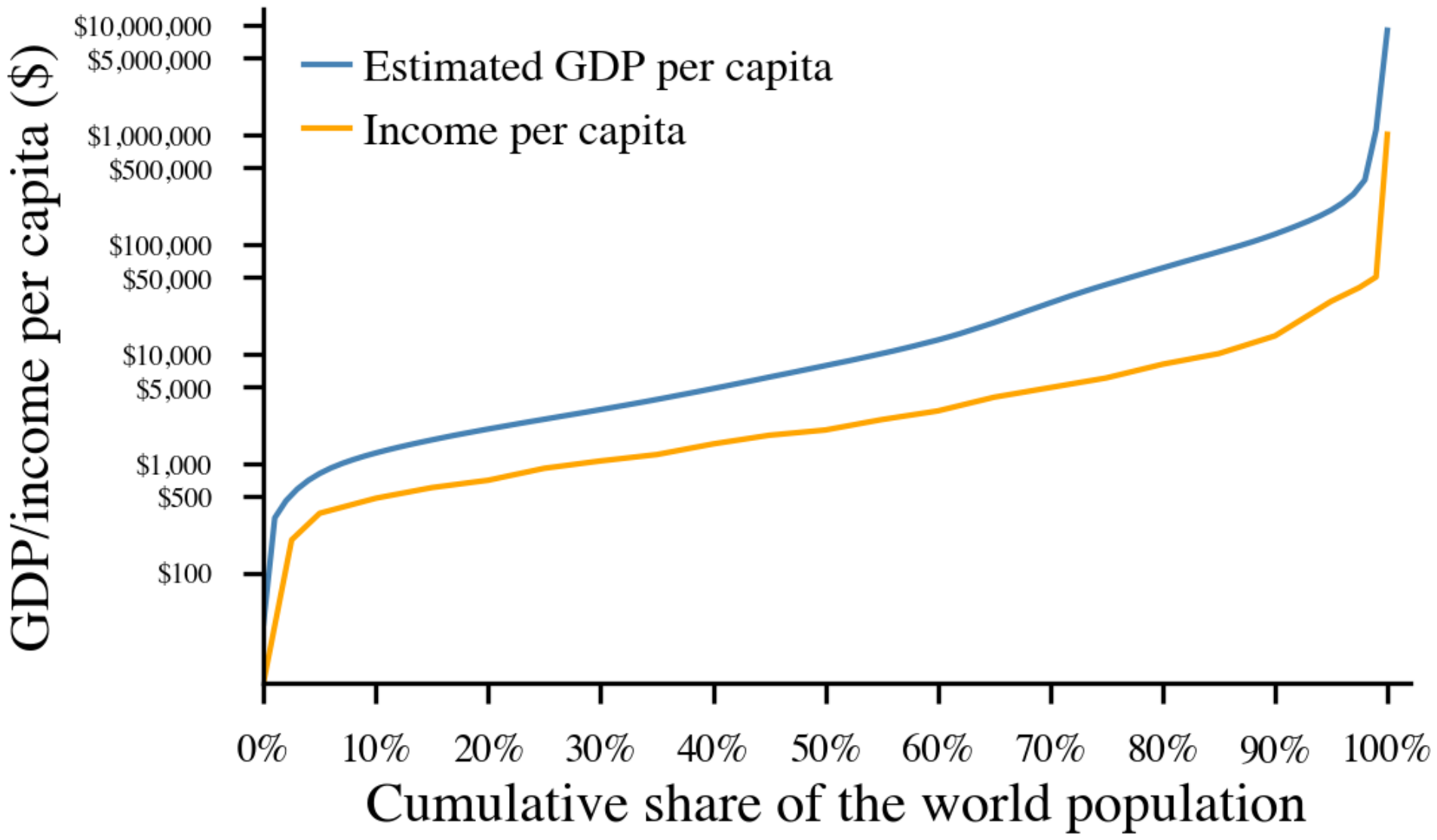

**Fig. S15 | The global income and estimated wealth distribution.** Orange line shows the global income distribution in 2013, based on household income surveys for more than a hundred countries. Blue line shows the distribution of predicted "absolute wealth", a measure of per capita GDP, which is derived from the Relative Wealth Index that is the focus of this paper.

| | Country | Code | Survey Year | # households | # villages |
|---|---|---|---|---|---|
| *1* | Albania | AL | 2017-18 | 15,823 | 715 |
| *2* | Angola | AO | 2015-16 | 16,109 | 625 |
| *3* | Armenia | AM | 2015-16 | 7,893 | 313 |
| *4* | Bangladesh | BD | 2014 | 17,270 | 599 |
| *5* | Benin | BJ | 2017-18 | 13,776 | 540 |
| *6* | Bolivia | BO | 2008 | 19,526 | 998 |
| *7* | Burkina Faso | BF | 2010 | 13,617 | 541 |
| *8* | Burundi | BI | 2016-17 | 15,921 | 552 |
| *9* | Cambodia | KH | 2014 | 15,825 | 611 |
| *10* | Cameroon | CM | 2011 | 14,189 | 577 |
| *11* | Chad | TD | 2014-15 | 17,233 | 624 |
| *12* | Colombia | CO | 2010 | 50,218 | 4,868 |
| *13* | Congo (DRC) | CD | 2013-14 | 16,680 | 492 |
| *14* | Cote d'Ivoire | CI | 2011-12 | 9,394 | 341 |
| *15* | Dominican Republic | DO | 2013 | 11,464 | 524 |
| *16* | Egypt | EG | 2014 | 27,904 | 1,817 |
| *17* | eSwatini / Swaziland | SZ | 2006-07 | 4,756 | 270 |
| *18* | Ethiopia | ET | 2016 | 16,157 | 622 |
| *19* | Gabon | GA | 2012 | 9,638 | 332 |
| *20* | Ghana | GH | 2014 | 11,716 | 423 |
| *21* | Guatemala | GT | 2014-15 | 21,263 | 853 |
| *22* | Guinea | GN | 2018 | 7,912 | 401 |
| *23* | Guyana | GY | 2009 | 5,418 | 312 |
| *24* | Haiti | HT | 2016-17 | 13,405 | 450 |
| *25* | Honduras | HN | 2011-12 | 20,985 | 1,128 |
| *26* | India | IN | 2015-16 | 598,733 | 28,393 |
| *27* | Indonesia | ID | 2002-03 | 31,393 | 1,319 |
| *28* | Jordan | JO | 2012 | 15,190 | 806 |
| *29* | Kenya | KE | 2014 | 36,224 | 1,585 |
| *30* | Kyrgyz Republic | KG | 2012 | 7,989 | 314 |
| *31* | Lesotho | LS | 2014 | 9,402 | 399 |
| *32* | Liberia | LR | 2013 | 9,333 | 322 |
| *33* | Madagascar | MG | 2008-09 | 17,578 | 585 |
| *34* | Malawi | MW | 2015-16 | 26,361 | 850 |
| *35* | Mali | ML | 2018 | 8,918 | 328 |
| *36* | Moldova | MD | 2005 | 11,066 | 399 |
| *37* | Morocco | MA | 2003-04 | 11,513 | 480 |
| *38* | Mozambique | MZ | 2011 | 13,899 | 609 |
| *39* | Myanmar | MM | 2015-16 | 12,500 | 441 |
| *40* | Namibia | NA | 2013 | 9,849 | 550 |
| *41* | Nepal | NP | 2016 | 11,040 | 383 |
| *42* | Nigeria | NG | 2018 | 39,540 | 1,359 |
| *43* | Pakistan | PK | 2017-18 | 14,517 | 560 |
| *44* | Peru | PE | 2009 | 26,809 | 1,131 |
| *45* | Philippines | PH | 2017 | 26,673 | 1,213 |
| *46* | Rwanda | RW | 2014-15 | 12,699 | 492 |
| *47* | Senegal | SN | 2016 | 4,437 | 214 |
| *48* | Sierra Leone | SL | 2013 | 12,629 | 435 |
| *49* | South Africa | ZA | 2016 | 11,083 | 746 |
| *50* | Tajikistan | TJ | 2017 | 7,821 | 365 |
| *51* | Tanzania | TZ | 2015-16 | 12,563 | 608 |
| *52* | Timor-Leste | TL | 2016 | 11,502 | 455 |
| *53* | Togo | TG | 2013-14 | 9,549 | 330 |
| *54* | Uganda | UG | 2016 | 19,284 | 685 |
| *55* | Zambia | ZM | 2018 | 12,595 | 535 |
| *56* | Zimbabwe | ZW | 2015 | 10,534 | 400 |
| | ***Total*** | | | ***1,457,315*** | ***66,819*** |

**Table S1 | Ground truth data**. The relative wealth prediction model is trained on nationally representative Demographic and Health Surveys from 56 countries. See www.dhsprogram.com.

| | Resolution | Source | Min | Mean | Median | Max |
|---|---|---|---|---|---|---|
| *INPUT DATA* | | | | | | |
| **Road density*** | lat/lon | Open Street Map[1] | 0 | 0.0007 | 0 | 0.08 |
| **Urban or built up*** | 15 arc-seconds | NASA (MODIS)[2] | | | | |
| **Elevation*** | 3 arc-seconds (~ 90 meters) | USGS[3] | -24 | 612 | 294 | 7643 |
| **Slope*** | 3 arc-seconds (~ 90 meters) | USGS[3] | 0.0 | 0.024 | 0.008 | 2.108 |
| **Precipitation*** | 0.25-degree | NASA/Japan Aerospace Exploration Agency[4] | -48.7 | 1.7 | 0.0 | 2233.6 |
| **Population*** | 1 arc-second (~ 30m) | Humanitarian Data Exchange[5] | 10 | 608 | 73 | 516163 |
| **# Cell towers⁺** | 2.4km tiles | Facebook[6] | 0 | 13 | 0 | 71004 |
| **# WiFi access points⁺** | 2.4km tiles | Facebook[6] | 0 | 369 | 1 | 1949963 |
| **# Mobile devices⁺** | 2.4km tiles | Facebook[6] | 0 | 217 | 0 | 454962 |
| **# Android devices⁺** | 2.4km tiles | Facebook[6] | 0 | 168 | 0 | 291831 |
| **# iOS devices⁺** | 2.4km tiles | Facebook[6] | 0 | 49 | 0 | 204058 |
| **Nightlights / Radiance (VIIRS)*** | 15 arc-seconds | National Centers for Environmental Information Earth Observation Group[7] | 0.0 | 1.9 | 0.3 | 58843.3 |
| **Satellite Imagery⁺** | 0.58 m/pixel | Digital Globe[8] (Bing tile 15) | | | | |
| *GROUND TRUTH DATA* | | | | | | |
| **Household survey*** | cluster | Demographic and Health Surveys[9] | | | | |
| **Census data*** | level-2 admin | IPUMS[10] | | | | |
| **Regional GDPpc** | TL3 regions | OECD[11] | | | | |
| **Togo** | GPS coordinates | Government of Togo | | | | |
| **PPI data⁺** | GPS coordinates | GiveDirectly[12] | | | | |
| **GDP and Gini*** | country | Multiple - see Table S6 | | | | |

**Table S2 | Data Sources**. Summary statistics for the different datasets that are used as input to the machine learning algorithms. We use the most recently available data layer from each source. Publicly available data denoted by *; Data requiring license or other restrictions denoted by ⁺.

*Sources:*
[1] http://www.openstreetmap.org
[2] http://www.landcover.org/data/lc/
[3] https://lta.cr.usgs.gov/SRTM1Arc
[4] https://disc.gsfc.nasa.gov/
[5] https://data.humdata.org/dataset/highresolutionpopulationdensitymaps
[6] https://research.fb.com/category/connectivity/
[7] https://www.ngdc.noaa.gov/eog
[8] http://www.digitalglobe.com/
[9] http://www.dhsprogram.com/
[10] https://international.ipums.org/international/
[11] https://stats.oecd.org/Index.aspx?DataSetCode=PDB_LV
[12] http://www.givedirectly.org/

| | Country | Survey Year | # Households | # Individuals | # Admin units | $R^2$ |
|---|---|---|---|---|---|---|
| *1* | Benin | 2013 | 180,621 | 1,009,693 | 77 | 0.90 |
| *2* | Dominican Republic | 2010 | 268,637 | 943,784 | 67 | 0.90 |
| *3* | Haiti | 2003 | 179,190 | 838,045 | 28 | 0.96 |
| *4* | Honduras | 2001 | 123,584 | 608,620 | 99 | 0.93 |
| *5* | Lesotho | 2006 | 41,726 | 180,208 | 64 | 0.77 |
| *6* | Mexico | 2015 | 2,927,196 | 11,344,365 | 32 | 0.80 |
| *7* | Nicaragua | 2005 | 105,629 | 515,485 | 70 | 0.94 |
| *8* | Panama | 2010 | 95,579 | 341,118 | 36 | 0.90 |
| *9* | Rwanda | 2012 | 242,461 | 1,038,369 | 30 | 0.82 |
| *10* | Senegal | 2002 | 107,999 | 994,562 | 28 | 0.95 |
| *11* | Sierra Leone | 2004 | 82,518 | 494,298 | 108 | 0.83 |
| *12* | Tanzania | 2012 | 950,776 | 4,498,022 | 114 | 0.82 |
| *13* | Togo | 2010 | 121,237 | 584,859 | 37 | 0.93 |
| *14* | Venezuela | 2001 | 543,475 | 2,306,489 | 158 | 0.54 |
| *15* | Zambia | 2010 | 250,805 | 1,321,973 | 55 | 0.94 |
| | ***Total*** | | ***6,221,433*** | ***27,019,890*** | ***1,003*** | ***Avg: 0.86*** |

**Table S3 | Census validation data**. Census data from 27 million individuals in 15 countries were used to provide independent validation of the wealth estimates. Data obtained from IPUMS (*41*). The final column indicates the proportion of variance in wealth (as measured in the census) explained by the model's wealth estimates (RWI) – see Fig. S4.

| | (1) Base specification | | (2) Incl. imagery features | | (3) Excluding all RWI features | |
|---|---|---|---|---|---|---|
| | *β* | SE | *β* | SE | *β* | SE |
| ln(Dist. to closest DHS country) | 0.0846*** | 0.007 | 0.0922*** | 0.008 | 0.0761*** | 0.007 |
| ln(Dist. the closest DHS cluster) | 0.0217** | 0.01 | 0.0153 | 0.01 | 0.0206** | 0.01 |
| ln(# neighbor countries w/ DHS) | 0.0124** | 0.005 | 0.0236*** | 0.006 | 0 | 0.005 |
| ln(# DHS clusters w/in 50 km) | -0.0115*** | 0.003 | -0.0084*** | 0.003 | -0.0158*** | 0.003 |
| ln(# DHS clusters w/in 250 km) | -0.0131*** | 0.002 | -0.0144*** | 0.002 | -0.0106*** | 0.002 |
| ln(# DHS clusters w/in 500 km) | -0.0004 | 0.001 | 0.0025 | 0.002 | -0.0013 | 0.001 |
| ln(# DHS clusters w/in 1000 km) | 0.0225*** | 0.002 | 0.0231*** | 0.002 | 0.023*** | 0.002 |
| Is island | 0.0209** | 0.01 | 0.0449*** | 0.011 | -0.0038 | 0.01 |
| Is landlocked | -0.0153** | 0.006 | -0.0638*** | 0.007 | -0.0155*** | 0.006 |
| Is America | -0.1066*** | 0.01 | -0.1101*** | 0.012 | -0.0934*** | 0.009 |
| Is Asia | 0.071*** | 0.007 | -0.0025 | 0.009 | 0.0825*** | 0.007 |
| Is Europe | -0.041*** | 0.015 | -0.129*** | 0.018 | -0.0664*** | 0.014 |
| ln(Area) | -0.0349*** | 0.004 | -0.0407*** | 0.004 | -0.034*** | 0.003 |
| ln(Country population) | 0.0044* | 0.003 | 0.0187*** | 0.003 | 0.011*** | 0.003 |
| ln(GDP per capita) | -0.0151*** | 0.004 | -0.0313*** | 0.005 | 0.0039 | 0.004 |
| Gini | 0.5347*** | 0.04 | 0.1347*** | 0.047 | 0.578*** | 0.039 |
| Road density | -0.7382* | 0.382 | 0.8917* | 0.497 | | |
| ln(Slope) | -0.1535** | 0.07 | -0.2316*** | 0.082 | | |
| ln(Elevation) | 0.0144*** | 0.001 | 0.0087*** | 0.002 | | |
| ln(Precipitation) | 0.0166*** | 0.004 | 0.0147*** | 0.004 | | |
| Is urban or built up | -0.0751*** | 0.006 | -0.0685*** | 0.007 | | |
| ln(Radiance) | -0.0033 | 0.004 | -0.0195*** | 0.004 | | |
| ln(Tile population) | 0.012*** | 0.002 | 0.0112*** | 0.002 | | |
| ln(# cell towers) | 0.0022 | 0.004 | -0.0015 | 0.004 | | |
| ln(# Wifi access points) | 0.0227*** | 0.003 | 0.0208*** | 0.003 | | |
| ln(# mobile devices) | 0.1581*** | 0.02 | 0.1696*** | 0.021 | | |
| ln(# Android devices) | -0.1423*** | 0.02 | -0.1493*** | 0.02 | | |
| ln(# iOS devices) | -0.0304*** | 0.003 | -0.0308*** | 0.003 | | |
| Satellite image features? | No | | Yes | | No | |
| Constant | -0.225*** | 0.059 | -0.0905 | 0.069 | -0.1614*** | 0.058 |

**Table S4 | Correlates of model error**. Table shows coefficients and standard errors from a regression of model error (defined as the absolute value of the village-level residual from the predictive model) on a set of characteristics of the village. Three columns correspond to three different model specifications. Data sources for country-level characteristics are provided in Table S6. *significant at 10 percent; ** significant at 5 percent; *** significant at 1 percent.

| | | | Estimated error | | | Mean squared error | | |
|---|---|---|---|---|---|---|---|---|
| | Country | Code | Mean | Median | S.D. | Mean | Median | S.D. |
| 1 | Afghanistan | AF | 0.217 | 0.213 | 0.024 | | | |
| 2 | Albania | AL | 0.440 | 0.435 | 0.061 | 0.551 | 0.203 | 1.257 |
| 3 | Algeria | DZ | 0.325 | 0.308 | 0.060 | | | |
| 4 | American Samoa | AS | 0.647 | 0.673 | 0.057 | | | |
| 5 | Angola | AO | 0.373 | 0.368 | 0.029 | 0.390 | 0.227 | 0.524 |
| 6 | Argentina | AR | 0.400 | 0.394 | 0.052 | | | |
| 7 | Armenia | AM | 0.427 | 0.408 | 0.049 | 0.467 | 0.242 | 0.634 |
| 8 | Azerbaijan | AZ | 0.260 | 0.240 | 0.053 | | | |
| 9 | Bangladesh | BD | 0.494 | 0.503 | 0.051 | 0.414 | 0.264 | 0.467 |
| 10 | Belarus | BY | 0.288 | 0.273 | 0.039 | | | |
| 11 | Belize | BZ | 0.378 | 0.359 | 0.049 | | | |
| 12 | Benin | BJ | 0.316 | 0.305 | 0.032 | 0.282 | 0.113 | 0.444 |
| 13 | Bhutan | BT | 0.378 | 0.365 | 0.040 | | | |
| 14 | Bolivia | BO | 0.396 | 0.387 | 0.035 | 0.472 | 0.256 | 0.610 |
| 15 | Bosnia & Herzegovina | BA | 0.360 | 0.351 | 0.046 | | | |
| 16 | Botswana | BW | 0.393 | 0.377 | 0.036 | | | |
| 17 | Brazil | BR | 0.395 | 0.392 | 0.047 | | | |
| 18 | Bulgaria | BG | 0.409 | 0.407 | 0.046 | | | |
| 19 | Burkina Faso | BF | 0.301 | 0.294 | 0.026 | 0.258 | 0.062 | 0.718 |
| 20 | Burundi | BI | 0.269 | 0.259 | 0.030 | 0.429 | 0.169 | 0.805 |
| 21 | Cabo Verde | CV | 0.532 | 0.528 | 0.062 | | | |
| 22 | Cambodia | KH | 0.503 | 0.497 | 0.061 | 0.387 | 0.191 | 0.559 |
| 23 | Cameroon | CM | 0.364 | 0.355 | 0.033 | 0.372 | 0.202 | 0.445 |
| 24 | Central African Republic | CF | 0.382 | 0.380 | 0.013 | | | |
| 25 | Chad | TD | 0.346 | 0.348 | 0.023 | 0.335 | 0.067 | 0.797 |
| 26 | China | CN | 0.372 | 0.349 | 0.064 | | | |
| 27 | Colombia | CO | 0.476 | 0.465 | 0.054 | 0.659 | 0.209 | 1.152 |
| 28 | Comoros | KM | 0.537 | 0.510 | 0.055 | | | |
| 29 | Congo | CG | 0.335 | 0.333 | 0.013 | | | |
| 30 | Congo, Dem. Rep. | CD | 0.378 | 0.375 | 0.019 | 0.307 | 0.082 | 0.612 |
| 31 | Costa Rica | CR | 0.530 | 0.537 | 0.055 | | | |
| 32 | Cote d'Ivoire | CI | 0.361 | 0.348 | 0.040 | 0.404 | 0.210 | 0.593 |
| 33 | Cuba | CU | 0.412 | 0.398 | 0.045 | | | |
| 34 | Djibouti | DJ | 0.341 | 0.336 | 0.024 | | | |
| 35 | Dominica | DM | 0.573 | 0.592 | 0.057 | | | |
| 36 | Dominican Republic | DO | 0.430 | 0.434 | 0.049 | 0.709 | 0.378 | 0.890 |
| 37 | Ecuador | EC | 0.443 | 0.429 | 0.051 | | | |
| 38 | Egypt | EG | 0.455 | 0.476 | 0.077 | 0.466 | 0.252 | 0.585 |
| 39 | El Salvador | SV | 0.417 | 0.420 | 0.038 | | | |
| 40 | Equatorial Guinea | GQ | 0.335 | 0.325 | 0.031 | | | |
| 41 | Eritrea | ER | 0.279 | 0.276 | 0.012 | | | |
| 42 | Eswatini | SZ | 0.469 | 0.454 | 0.044 | 0.466 | 0.309 | 0.521 |
| 43 | Ethiopia | ET | 0.359 | 0.360 | 0.029 | 0.236 | 0.075 | 0.427 |
| 44 | Fiji | FJ | 0.502 | 0.481 | 0.050 | | | |
| 45 | Gabon | GA | 0.390 | 0.387 | 0.021 | 0.347 | 0.182 | 0.418 |
| 46 | Gambia | GM | 0.291 | 0.271 | 0.045 | | | |
| 47 | Georgia | GE | 0.334 | 0.315 | 0.048 | | | |
| 48 | Ghana | GH | 0.348 | 0.330 | 0.047 | 0.363 | 0.180 | 0.567 |
| 49 | Grenada | GD | 0.583 | 0.591 | 0.046 | | | |
| 50 | Guatemala | GT | 0.460 | 0.456 | 0.054 | 0.553 | 0.311 | 0.704 |
| 51 | Guinea | GN | 0.250 | 0.241 | 0.029 | 0.258 | 0.117 | 0.459 |
| 52 | Guinea-Bissau | GW | 0.352 | 0.346 | 0.020 | | | |
| 53 | Guyana | GY | 0.412 | 0.399 | 0.039 | 0.388 | 0.192 | 0.548 |
| 54 | Haiti | HT | 0.338 | 0.326 | 0.043 | 0.318 | 0.127 | 0.492 |
| 55 | Honduras | HN | 0.437 | 0.420 | 0.049 | 0.401 | 0.226 | 0.529 |
| 56 | India | IN | 0.465 | 0.466 | 0.053 | 0.506 | 0.264 | 0.651 |
| 57 | Indonesia | ID | 0.438 | 0.431 | 0.068 | 0.580 | 0.300 | 0.767 |
| 58 | Iran, Islamic Rep. | IR | 0.368 | 0.363 | 0.042 | | | |
| 59 | Iraq | IQ | 0.374 | 0.347 | 0.064 | | | |
| 60 | Jamaica | JM | 0.543 | 0.553 | 0.044 | | | |
| 61 | Jordan | JO | 0.459 | 0.449 | 0.060 | 0.801 | 0.530 | 0.864 |
| 62 | Kazakhstan | KZ | 0.288 | 0.277 | 0.061 | | | |
| 63 | Kenya | KE | 0.391 | 0.371 | 0.047 | 0.363 | 0.139 | 0.602 |
| 64 | Kiribati | KI | 0.523 | 0.499 | 0.049 | | | |
| 65 | Korea, Dem. People's Republic | KP | 0.607 | 0.602 | 0.012 | | | |
| 66 | Kyrgyzstan | KG | 0.255 | 0.248 | 0.029 | 0.364 | 0.161 | 0.519 |
| 67 | Lao People's Dem. Republic | LA | 0.376 | 0.359 | 0.048 | | | |
| 68 | Lebanon | LB | 0.470 | 0.480 | 0.042 | | | |
| 69 | Lesotho | LS | 0.448 | 0.439 | 0.035 | 0.300 | 0.171 | 0.414 |
| 70 | Liberia | LR | 0.284 | 0.275 | 0.030 | 0.239 | 0.097 | 0.402 |
| 71 | Libya | LY | 0.281 | 0.259 | 0.050 | | | |
| 72 | Madagascar | MG | 0.359 | 0.360 | 0.026 | 0.288 | 0.100 | 0.593 |
| 73 | Malawi | MW | 0.378 | 0.366 | 0.034 | 0.401 | 0.132 | 0.756 |
| 74 | Malaysia | MY | 0.505 | 0.513 | 0.047 | | | |
| 75 | Maldives | MV | 0.671 | 0.669 | 0.059 | | | |
| 76 | Mali | ML | 0.274 | 0.269 | 0.025 | 0.206 | 0.065 | 0.387 |
| 77 | Marshall Islands | MH | 0.566 | 0.552 | 0.040 | | | |
| 78 | Mauritania | MR | 0.270 | 0.264 | 0.025 | | | |
| 79 | Mauritius | MU | 0.582 | 0.586 | 0.037 | | | |
| 80 | Mexico | MX | 0.447 | 0.445 | 0.065 | | | |
| 81 | Micronesia, Fed. Sts. | FM | 0.537 | 0.510 | 0.047 | | | |
| 82 | Moldova | MD | 0.403 | 0.396 | 0.049 | 0.565 | 0.268 | 0.735 |
| 83 | Mongolia | MN | 0.290 | 0.268 | 0.049 | | | |
| 84 | Montenegro | ME | 0.288 | 0.270 | 0.049 | | | |
| 85 | Morocco | MA | 0.437 | 0.420 | 0.050 | 0.640 | 0.298 | 0.792 |
| 86 | Mozambique | MZ | 0.372 | 0.366 | 0.027 | 0.300 | 0.138 | 0.510 |
| 87 | Myanmar | MM | 0.405 | 0.389 | 0.058 | 0.374 | 0.180 | 0.532 |
| 88 | Namibia | NA | 0.466 | 0.452 | 0.038 | 0.526 | 0.267 | 0.699 |
| 89 | Nauru | NR | 0.686 | 0.690 | 0.045 | | | |
| 90 | Nepal | NP | 0.425 | 0.414 | 0.052 | 0.378 | 0.217 | 0.473 |
| 91 | Nicaragua | NI | 0.342 | 0.326 | 0.044 | | | |
| 92 | Niger | NE | 0.275 | 0.270 | 0.020 | | | |
| 93 | Nigeria | NG | 0.385 | 0.373 | 0.044 | 0.406 | 0.189 | 0.597 |
| 94 | North Macedonia | MK | 0.273 | 0.259 | 0.047 | | | |
| 95 | Pakistan | PK | 0.430 | 0.420 | 0.052 | 0.550 | 0.270 | 0.753 |
| 96 | Papua New Guinea | PG | 0.339 | 0.333 | 0.024 | | | |
| 97 | Paraguay | PY | 0.416 | 0.403 | 0.044 | | | |
| 98 | Peru | PE | 0.438 | 0.423 | 0.047 | 0.430 | 0.212 | 0.573 |
| 99 | Philippines | PH | 0.514 | 0.516 | 0.057 | 0.487 | 0.232 | 0.645 |
| 100 | Romania | RO | 0.364 | 0.369 | 0.043 | | | |
| 101 | Russian Federation | RU | 0.346 | 0.318 | 0.050 | | | |
| 102 | Rwanda | RW | 0.343 | 0.331 | 0.044 | 0.395 | 0.159 | 0.622 |
| 103 | Saint Lucia | LC | 0.644 | 0.652 | 0.057 | | | |
| *104* | Samoa | WS | 0.547 | 0.523 | 0.056 | | | |
| *105* | Sao Tome & Principe | ST | 0.445 | 0.417 | 0.060 | | | |
| *106* | Senegal | SN | 0.401 | 0.386 | 0.039 | 0.279 | 0.133 | 0.410 |
| *107* | Serbia | RS | 0.321 | 0.319 | 0.046 | | | |
| *108* | Sierra Leone | SL | 0.274 | 0.265 | 0.029 | 0.211 | 0.078 | 0.626 |
| *109* | Solomon Islands | SB | 0.399 | 0.395 | 0.020 | | | |
| *110* | Somalia | SO | 0.301 | 0.299 | 0.018 | | | |
| *111* | South Africa | ZA | 0.514 | 0.509 | 0.055 | 0.834 | 0.292 | 1.402 |
| *112* | Sri Lanka | LK | 0.567 | 0.572 | 0.039 | | | |
| *113* | St. Vincent and the Grenadines | VC | 0.629 | 0.646 | 0.059 | | | |
| *114* | Sudan | SD | 0.413 | 0.405 | 0.028 | | | |
| *115* | Suriname | SR | 0.359 | 0.341 | 0.040 | | | |
| *116* | Syrian Arab Republic | SY | 0.375 | 0.355 | 0.052 | | | |
| *117* | Tajikistan | TJ | 0.296 | 0.289 | 0.029 | 0.509 | 0.266 | 0.596 |
| *118* | Tanzania | TZ | 0.331 | 0.325 | 0.032 | 0.315 | 0.162 | 0.442 |
| *119* | Thailand | TH | 0.441 | 0.442 | 0.031 | | | |
| *120* | Timor-Leste | TL | 0.416 | 0.395 | 0.046 | 0.308 | 0.132 | 0.455 |
| *121* | Togo | TG | 0.295 | 0.284 | 0.031 | 0.196 | 0.093 | 0.267 |
| *122* | Tonga | TO | 0.571 | 0.533 | 0.061 | | | |
| *123* | Tunisia | TN | 0.404 | 0.396 | 0.056 | | | |
| *124* | Turkey | TR | 0.480 | 0.479 | 0.054 | | | |
| *125* | Turkmenistan | TM | 0.375 | 0.371 | 0.017 | | | |
| *126* | Tuvalu | TV | 0.595 | 0.563 | 0.052 | | | |
| *127* | Uganda | UG | 0.356 | 0.345 | 0.034 | 0.285 | 0.132 | 0.413 |
| *128* | Ukraine | UA | 0.223 | 0.219 | 0.047 | | | |
| *129* | Uzbekistan | UZ | 0.322 | 0.310 | 0.039 | | | |
| *130* | Vanuatu | VU | 0.433 | 0.420 | 0.032 | | | |
| *131* | Venezuela, RB | VE | 0.421 | 0.410 | 0.037 | | | |
| *132* | Viet Nam | VN | 0.399 | 0.402 | 0.038 | | | |
| *133* | West Bank and Gaza | PS | 0.468 | 0.470 | 0.042 | | | |
| *134* | Yemen | YE | 0.351 | 0.343 | 0.041 | | | |
| *135* | Zambia | ZM | 0.374 | 0.369 | 0.025 | 0.318 | 0.138 | 0.602 |
| *136* | Zimbabwe | ZW | 0.365 | 0.357 | 0.031 | 0.304 | 0.109 | 0.496 |

**Table S5 | Estimates of model error in all low and middle-income countries.** Table indicates mean, median, and standard deviation of predicted model error in all LMICs (columns 3-5). In countries where ground truth DHS data exist, table reports mean, median, and standard deviation of mean squared prediction error (columns 6-8).

| | Country | Code | GDP per capita | Year (GDP) | Source (GDP per capita) | Gini | Year (Gini) | Source (Gini) |
|---|---|---|---|---|---|---|---|---|
| 1 | Afghanistan | AF | 520.90 | 2018 | World Bank[1] | 0.278 | - | [6] |
| 2 | Albania | AL | 5253.63 | 2018 | World Bank[1] | 0.29 | 2012 | World Bank[2] |
| 3 | Algeria | DZ | 4278.85 | 2018 | World Bank[1] | 0.276 | 2011 | World Bank[2] |
| 4 | American Samoa | AS | 11398.78 | 2017 | World Bank[1] | 0.387 | - | WS[4] |
| 5 | Angola | AO | 3432.39 | 2018 | World Bank[1] | 0.427 | 2008 | World Bank[2] |
| 6 | Argentina | AR | 11652.57 | 2018 | World Bank[1] | 0.406 | 2017 | World Bank[2] |
| 7 | Armenia | AM | 4212.07 | 2018 | World Bank[1] | 0.336 | 2017 | World Bank[2] |
| 8 | Azerbaijan | AZ | 4721.18 | 2018 | World Bank[1] | 0.266 | 2005 | World Bank[2] |
| 9 | Bangladesh | BD | 1698.26 | 2018 | World Bank[1] | 0.324 | 2016 | World Bank[2] |
| 10 | Belarus | BY | 6289.94 | 2018 | World Bank[1] | 0.254 | 2017 | World Bank[2] |
| 11 | Belize | BZ | 5025.18 | 2018 | World Bank[1] | 0.533 | 1999 | World Bank[2] |
| 12 | Benin | BJ | 901.95 | 2018 | World Bank[1] | 0.478 | 2015 | World Bank[2] |
| 13 | Bhutan | BT | 3360.27 | 2018 | World Bank[1] | 0.374 | 2017 | World Bank[2] |
| 14 | Bolivia | BO | 3548.59 | 2018 | World Bank[1] | 0.44 | 2017 | World Bank[2] |
| 15 | Bosnia & Herzegovina | BA | 5951.32 | 2018 | World Bank[1] | 0.33 | 2011 | World Bank[2] |
| 16 | Botswana | BW | 8258.64 | 2018 | World Bank[1] | 0.533 | 2015 | World Bank[2] |
| 17 | Brazil | BR | 8920.76 | 2018 | World Bank[1] | 0.533 | 2017 | World Bank[2] |
| 18 | Bulgaria | BG | 9272.63 | 2018 | World Bank[1] | 0.374 | 2014 | World Bank[2] |
| 19 | Burkina Faso | BF | 731.17 | 2018 | World Bank[1] | 0.353 | 2014 | World Bank[2] |
| 20 | Burundi | BI | 275.43 | 2018 | World Bank[1] | 0.386 | 2013 | World Bank[2] |
| 21 | Cabo Verde | CV | 3654.01 | 2018 | World Bank[1] | 0.472 | 2007 | World Bank[2] |
| 22 | Cambodia | KH | 1512.13 | 2018 | World Bank[1] | 0.379 | - | [5] |
| 23 | Cameroon | CM | 1526.88 | 2018 | World Bank[1] | 0.466 | 2014 | World Bank[2] |
| 24 | Central African Republic | CF | 509.97 | 2018 | World Bank[1] | 0.562 | 2008 | World Bank[2] |
| 25 | Chad | TD | 730.24 | 2018 | World Bank[1] | 0.433 | 2011 | World Bank[2] |
| 26 | China | CN | 9770.85 | 2018 | World Bank[1] | 0.386 | 2015 | World Bank[2] |
| 27 | Colombia | CO | 6651.29 | 2018 | World Bank[1] | 0.497 | 2017 | World Bank[2] |
| 28 | Comoros | KM | 1445.45 | 2018 | World Bank[1] | 0.453 | 2013 | World Bank[2] |
| 29 | Congo, Dem. Rep. | CD | 561.78 | 2018 | World Bank[1] | 0.421 | 2012 | World Bank[2] |
| 30 | Congo, Rep. | CG | 2147.77 | 2018 | World Bank[1] | 0.489 | 2011 | World Bank[2] |
| 31 | Costa Rica | CR | 12026.55 | 2018 | World Bank[1] | 0.483 | 2017 | World Bank[2] |
| 32 | Cote d'Ivoire | CI | 1715.53 | 2018 | World Bank[1] | 0.415 | 2015 | World Bank[2] |
| 33 | Cuba | CU | 8541.21 | 2017 | World Bank[1] | 0.38 | 2003 | [7] |
| 34 | Djibouti | DJ | 2050.20 | 2018 | World Bank[1] | 0.416 | 2017 | World Bank[2] |
| 35 | Dominica | DM | 7031.71 | 2018 | World Bank[1] | 0.47 | 2013 | [8] |
| 36 | Dominican Republic | DO | 7650.07 | 2018 | World Bank[1] | 0.457 | 2016 | World Bank[2] |
| 37 | Ecuador | EC | 6344.87 | 2018 | World Bank[1] | 0.447 | 2017 | World Bank[2] |
| 38 | Egypt, Arab Rep. | EG | 2549.14 | 2018 | World Bank[1] | 0.318 | 2015 | World Bank[2] |
| 39 | El Salvador | SV | 4058.24 | 2018 | World Bank[1] | 0.38 | 2017 | World Bank[2] |
| 40 | Equatorial Guinea | GQ | 10173.96 | 2018 | World Bank[1] | 0.38 | - | GA[4] |
| 41 | Eritrea | ER | 811.38 | 2011 | World Bank[1] | 0.292 | 2016 | [9] |
| 42 | Eswatini | SZ | 4139.96 | 2018 | World Bank[1] | 0.515 | 2009 | World Bank[2] |
| 43 | Ethiopia | ET | 772.31 | 2018 | World Bank[1] | 0.391 | 2015 | World Bank[2] |
| 44 | Fiji | FJ | 6202.16 | 2018 | World Bank[1] | 0.367 | 2013 | World Bank[2] |
| 45 | Gabon | GA | 8029.82 | 2018 | World Bank[1] | 0.38 | 2017 | World Bank[2] |
| 46 | Gambia, The | GM | 712.45 | 2018 | World Bank[1] | 0.359 | 2015 | World Bank[2] |
| 47 | Georgia | GE | 4344.63 | 2018 | World Bank[1] | 0.379 | 2017 | World Bank[2] |
| 48 | Ghana | GH | 2202.31 | 2018 | World Bank[1] | 0.435 | 2016 | World Bank[2] |
| 49 | Grenada | GD | 10833.66 | 2018 | World Bank[1] | 0.512 | - | LC[4] |
| 50 | Guatemala | GT | 4549.01 | 2018 | World Bank[1] | 0.483 | 2014 | World Bank[2] |
| 51 | Guinea | GN | 885.25 | 2018 | World Bank[1] | 0.337 | 2012 | World Bank[2] |
| 52 | Guinea-Bissau | GW | 777.97 | 2018 | World Bank[1] | 0.507 | 2010 | World Bank[2] |
| 53 | Guyana | GY | 4634.68 | 2018 | World Bank[1] | 0.446 | 1998 | World Bank[2] |
| 54 | Haiti | HT | 868.28 | 2018 | World Bank[1] | 0.411 | 2012 | World Bank[2] |
| 55 | Honduras | HN | 2482.73 | 2018 | World Bank[1] | 0.505 | 2017 | World Bank[2] |
| 56 | India | IN | 2015.59 | 2018 | World Bank[1] | 0.357 | 2011 | World Bank[2] |
| 57 | Indonesia | ID | 3893.60 | 2018 | World Bank[1] | 0.381 | 2017 | World Bank[2] |
| 58 | Iran, Islamic Rep. | IR | 5627.75 | 2017 | World Bank[1] | 0.4 | 2016 | World Bank[2] |
| 59 | Iraq | IQ | 5878.04 | 2018 | World Bank[1] | 0.295 | 2012 | World Bank[2] |
| 60 | Jamaica | JM | 5355.58 | 2018 | World Bank[1] | 0.455 | 2004 | World Bank[2] |
| 61 | Jordan | JO | 4247.77 | 2018 | World Bank[1] | 0.337 | 2010 | World Bank[2] |
| 62 | Kazakhstan | KZ | 9331.05 | 2018 | World Bank[1] | 0.275 | 2017 | World Bank[2] |
| 63 | Kenya | KE | 1710.51 | 2018 | World Bank[1] | 0.408 | 2015 | World Bank[2] |
| 64 | Kiribati | KI | 1625.29 | 2018 | World Bank[1] | 0.37 | 2006 | World Bank[2] |
| 65 | Korea, D.P.R. | KP | 1700.00 | 2015 | CIA[3] | 0.85 | 2004 | [12] |
| 66 | Kosovo | XK | 4281.29 | 2018 | World Bank[1] | 0.29 | 2017 | World Bank[2] |
| 67 | Kyrgyz Republic | KG | 1281.36 | 2018 | World Bank[1] | 0.273 | 2017 | World Bank[2] |
| 68 | Lao PDR | LA | 2567.54 | 2018 | World Bank[1] | 0.364 | 2012 | World Bank[2] |
| 69 | Lebanon | LB | 8269.79 | 2018 | World Bank[1] | 0.318 | 2011 | World Bank[2] |
| 70 | Lesotho | LS | 1324.28 | 2018 | World Bank[1] | 0.542 | 2010 | World Bank[2] |
| 71 | Liberia | LR | 674.21 | 2018 | World Bank[1] | 0.353 | 2016 | World Bank[2] |
| 72 | Libya | LY | 7235.03 | 2018 | World Bank[1] | 0.307 | - | [10] |
| 73 | Madagascar | MG | 460.75 | 2018 | World Bank[1] | 0.426 | 2012 | World Bank[2] |
| 74 | Malawi | MW | 389.40 | 2018 | World Bank[1] | 0.447 | 2016 | World Bank[2] |
| 75 | Malaysia | MY | 11238.96 | 2018 | World Bank[1] | 0.41 | 2015 | World Bank[2] |
| 76 | Maldives | MV | 10223.64 | 2018 | World Bank[1] | 0.384 | 2009 | World Bank[2] |
| 77 | Mali | ML | 901.40 | 2018 | World Bank[1] | 0.33 | 2009 | World Bank[2] |
| 78 | Marshall Islands | MH | 3621.17 | 2018 | World Bank[1] | 0.391 | - | TV[4] |
| 79 | Mauritania | MR | 1218.60 | 2018 | World Bank[1] | 0.326 | 2014 | World Bank[2] |
| 80 | Mauritius | MU | 11238.69 | 2018 | World Bank[1] | 0.358 | 2012 | World Bank[2] |
| 81 | Mexico | MX | 9698.08 | 2018 | World Bank[1] | 0.434 | 2016 | World Bank[2] |
| 82 | Micronesia, Fed. Sts. | FM | 3058.43 | 2018 | World Bank[1] | 0.401 | 2013 | World Bank[2] |
| 83 | Moldova | MD | 3189.36 | 2018 | World Bank[1] | 0.259 | 2017 | World Bank[2] |
| 84 | Mongolia | MN | 4103.70 | 2018 | World Bank[1] | 0.323 | 2016 | World Bank[2] |
| 85 | Montenegro | ME | 8760.69 | 2018 | World Bank[1] | 0.319 | 2014 | World Bank[2] |
| 86 | Morocco | MA | 3237.88 | 2018 | World Bank[1] | 0.395 | 2013 | World Bank[2] |
| 87 | Mozambique | MZ | 490.17 | 2018 | World Bank[1] | 0.54 | 2014 | World Bank[2] |
| 88 | Myanmar | MM | 1325.95 | 2018 | World Bank[1] | 0.381 | 2015 | World Bank[2] |
| 89 | Namibia | NA | 5931.45 | 2018 | World Bank[1] | 0.591 | 2015 | World Bank[2] |
| 90 | Nauru | NR | 9030.07 | 2018 | World Bank[1] | 0.371 | - | SB[4] |
| 91 | Nepal | NP | 1025.80 | 2018 | World Bank[1] | 0.328 | 2010 | World Bank[2] |
| 92 | Nicaragua | NI | 2028.90 | 2018 | World Bank[1] | 0.462 | 2014 | World Bank[2] |
| 93 | Niger | NE | 411.69 | 2018 | World Bank[1] | 0.343 | 2014 | World Bank[2] |
| 94 | Nigeria | NG | 2028.18 | 2018 | World Bank[1] | 0.43 | 2009 | World Bank[2] |
| 95 | North Macedonia | MK | 6083.72 | 2018 | World Bank[1] | 0.356 | 2015 | World Bank[2] |
| 96 | Pakistan | PK | 1472.89 | 2018 | World Bank[1] | 0.335 | 2015 | World Bank[2] |
| 97 | Papua New Guinea | PG | 2722.60 | 2018 | World Bank[1] | 0.419 | 2009 | World Bank[2] |
| 98 | Paraguay | PY | 5871.47 | 2018 | World Bank[1] | 0.488 | 2017 | World Bank[2] |
| 99 | Peru | PE | 6947.26 | 2018 | World Bank[1] | 0.433 | 2017 | World Bank[2] |
| 100 | Philippines | PH | 3102.71 | 2018 | World Bank[1] | 0.401 | 2015 | World Bank[2] |
| 101 | Romania | RO | 12301.19 | 2018 | World Bank[1] | 0.359 | 2015 | World Bank[2] |
| 102 | Russian Federation | RU | 11288.87 | 2018 | World Bank[1] | 0.377 | 2015 | World Bank[2] |
| 103 | Rwanda | RW | 772.97 | 2018 | World Bank[1] | 0.437 | 2016 | World Bank[2] |
| 104 | Samoa | WS | 4392.47 | 2018 | World Bank[1] | 0.387 | 2013 | World Bank[2] |
| 105 | Sao Tome & Principe | ST | 2001.14 | 2018 | World Bank[1] | 0.308 | 2010 | World Bank[2] |
| 106 | Senegal | SN | 1521.95 | 2018 | World Bank[1] | 0.403 | 2011 | World Bank[2] |
| 107 | Serbia | RS | 7234.00 | 2018 | World Bank[1] | 0.285 | 2015 | World Bank[2] |
| 108 | Sierra Leone | SL | 522.86 | 2018 | World Bank[1] | 0.34 | 2011 | World Bank[2] |
| 109 | Solomon Islands | SB | 2162.65 | 2018 | World Bank[1] | 0.371 | 2013 | World Bank[2] |
| 110 | Somalia | SO | 498.66 | 2018 | World Bank[1] | 0.397 | 2012 | [11] |
| 111 | South Africa | ZA | 6339.57 | 2018 | World Bank[1] | 0.63 | 2014 | World Bank[2] |
| 112 | South Sudan | SS | 283.49 | 2016 | World Bank[1] | 0.463 | 2009 | World Bank[2] |
| 113 | Sri Lanka | LK | 4102.48 | 2018 | World Bank[1] | 0.398 | 2016 | World Bank[2] |
| 114 | St. Lucia | LC | 10315.03 | 2018 | World Bank[1] | 0.512 | 2016 | World Bank[2] |
| 115 | St. Vincent & Grenadines | VC | 7377.68 | 2018 | World Bank[1] | 0.512 | - | LC[4] |
| 116 | Sudan | SD | 977.27 | 2018 | World Bank[1] | 0.354 | 2009 | World Bank[2] |
| 117 | Suriname | SR | 5950.21 | 2018 | World Bank[1] | 0.576 | 1999 | World Bank[2] |
| 118 | Syrian Arab Republic | SY | 2032.62 | 2007 | World Bank[1] | 0.358 | 2004 | World Bank[2] |
| 119 | Tajikistan | TJ | 826.62 | 2018 | World Bank[1] | 0.34 | 2015 | World Bank[2] |
| 120 | Tanzania | TZ | 1050.68 | 2018 | World Bank[1] | 0.378 | 2011 | World Bank[2] |
| 121 | Thailand | TH | 7273.56 | 2018 | World Bank[1] | 0.365 | 2017 | World Bank[2] |
| 122 | Timor-Leste | TL | 2035.53 | 2018 | World Bank[1] | 0.287 | 2014 | World Bank[2] |
| 123 | Togo | TG | 671.84 | 2018 | World Bank[1] | 0.431 | 2015 | World Bank[2] |
| 124 | Tonga | TO | 4364.02 | 2018 | World Bank[1] | 0.376 | 2015 | World Bank[2] |
| 125 | Tunisia | TN | 3446.61 | 2018 | World Bank[1] | 0.328 | 2015 | World Bank[2] |
| 126 | Turkey | TR | 9311.37 | 2018 | World Bank[1] | 0.419 | 2016 | World Bank[2] |
| 127 | Turkmenistan | TM | 6966.64 | 2018 | World Bank[1] | 0.408 | 1998 | World Bank[2] |
| 128 | Tuvalu | TV | 3700.71 | 2018 | World Bank[1] | 0.391 | 2010 | World Bank[2] |
| 129 | Uganda | UG | 643.14 | 2018 | World Bank[1] | 0.428 | 2016 | World Bank[2] |
| 130 | Ukraine | UA | 3095.17 | 2018 | World Bank[1] | 0.25 | 2016 | World Bank[2] |
| 131 | Uzbekistan | UZ | 1532.37 | 2018 | World Bank[1] | 0.353 | 2003 | World Bank[2] |
| 132 | Vanuatu | VU | 3033.41 | 2018 | World Bank[1] | 0.376 | 2010 | World Bank[2] |
| 133 | Venezuela, RB | VE | 16054.49 | 2014 | World Bank[1] | 0.469 | 2006 | World Bank[2] |
| 134 | Vietnam | VN | 2563.82 | 2018 | World Bank[1] | 0.353 | 2016 | World Bank[2] |
| 135 | West Bank and Gaza | PS | 3198.87 | 2018 | World Bank[1] | 0.337 | 2016 | World Bank[2] |
| 136 | Yemen, Rep. | YE | 944.41 | 2018 | World Bank[1] | 0.367 | 2014 | World Bank[2] |
| 137 | Zambia | ZM | 1539.90 | 2018 | World Bank[1] | 0.571 | 2015 | World Bank[2] |
| 138 | Zimbabwe | ZW | 2147.00 | 2018 | World Bank[1] | 0.432 | 2011 | World Bank[2] |

**Table S6 | Sources of country-level data**. While most of the country-level statistics come from the World Bank's Open Data portal, when the required indicators are missing we use data from the alternative data sources listed above. Sources below.

| | |
|---|---|
| 1 | https://data.worldbank.org/indicator/ny.gdp.pcap.cd |
| 2 | https://data.worldbank.org/indicator/SI.POV.GINI?locations=US-AF |
| 3 | https://www.cia.gov/library/publications/the-world-factbook/rankorder/2004rank.html |
| 4 | No Gini available. Gini from closest neighbor, based on orthodromic distance, is used instead (and country code is indicated in the table when applicable). |
| 5 | https://www.cia.gov/library/publications/the-world-factbook/rankorder/2172rank.html |
| 6 | http://hdr.undp.org/en/content/income-gini-coefficient |
| 7 | https://watermark.silverchair.com/bey026.pdf |
| 8 | https://www.indexmundi.com/dominica/distribution_of_family_income_gini_index.html |
| 9 | https://pdfs.semanticscholar.org/1f1d/6a9df57105dba86090b5904422af6f087b9a.pdf |
| 10 | http://www.ecineq.org/ecineq_nyc17/FILESx2017/CR2/p426.pdf |
| 11 | https://canada-vs-somalia.weebly.com/somalia.html |
| 12 | https://www.piie.com/blogs/north-korea-witness-transformation/distribution-income-north-korea |

| | (1) # of spatial units | (2) # of units with estimates | (3) % units with estimates | (4) # units with ground truth | (5) # units with estimates & ground truth | (6) # of households used to evaluate targeting | (7) $R^2$ | (8) Targeting accuracy, poorest 25% | (9) Targeting accuracy, poorest 50% | (10) Targeting Precision/Recall poorest 25% | (11) Targeting Precision/Recall poorest 50% |
|---|---|---|---|---|---|---|---|---|---|---|---|
| *Panel A: High-resolution estimates* | | | | | | | | | | | |
| Tiles | 10,187 | 10,187 | 100% | 923 | 923 | 6,149 | 0.60 | 0.73 | 0.79 | 0.47 | 0.79 |
| Canton targeting | 387 | 387 | 100% | 260 | 260 | 6149 | 0.56 | 0.73 | 0.77 | 0.47 | 0.77 |
| *Panel B: Imputation based on DHS data (not implementable due to incomplete coverage -- locations without DHS data are excluded* | | | | | | | | | | | |
| Prefecture targeting | 40 | 40 | 100% | 40 | 40 | 6,149 | 0.49 | 0.70 | 0.70 | 0.39 | 0.70 |
| Canton targeting | 387 | 185 | 47.8% | 260 | 149 | 4,509 | 0.52 | 0.76 | 0.80 | 0.53 | 0.80 |
| *Panel C: Imputation based on DHS data (imputing estimates for locations where no DHS data exist)* | | | | | | | | | | | |
| Prefecture targeting | 40 | 40 | 100% | 40 | 40 | 6,149 | 0.49 | 0.70 | 0.70 | 0.39 | 0.70 |
| Canton targeting | 387 | 387 | 100% | 260 | 260 | 6,149 | 0.53 | 0.72 | 0.75 | 0.44 | 0.75 |
| *Panel D: Imputation based on wealth of k-Nearest neighbors in DHS survey* | | | | | | | | | | | |
| Nearest Neighbor | - | - | - | - | - | - | 0.57 | 0.73 | 0.78 | 0.45 | 0.78 |
| Avg of 5 neighbors | - | - | - | - | - | - | 0.52 | 0.70 | 0.72 | 0.39 | 0.72 |

**Table S7 | Targeting simulations in Togo. a)** Panel A of table simulates the performance of an anti-poverty program that geographically targets households in the poorest 2.4km tiles in Togo, using the ML estimates of tile wealth. Panels B and C simulate the geo graphic targeting of households in the poorest prefectures and cantons of Togo, where the most recent DHS survey is used to estimate the average wealth of each administrative region. Panel B ignores households regions with no DHS surveys; Panel C assigns such households the average wealth of the geographic unit closest to the household. Panel D simulates targets poor households where the wealth of a household is inferred based on the average wealth of the households in the DHS cluster physically closest to it. Simulations are evaluated using the 2018-19 EHCVM survey.

| | (1) # of spatial units | (2) # of units with estimates | (3) % units with estimates | (4) # units with ground truth | (5) # units with estimates & ground truth | (6) # of households used to evaluate targeting | (7) $R^2$ | (8) Targeting accuracy, poorest 25% | (9) Targeting accuracy, poorest 50% | (10) Targeting Precision/Recall poorest 25% | (11) Targeting Precision/Recall poorest 50% |
|---|---|---|---|---|---|---|---|---|---|---|---|
| *Panel A: High-resolution estimates* | | | | | | | | | | | |
| Tile targeting | 159,147 | 159,147 | 100% | 2,446 | 2,446 | 22,060 | 0.53 | 0.79 | 0.79 | 0.57 | 0.79 |
| Ward targeting | 8,808 | 8,808 | 100% | 2,016 | 2,016 | 22060 | 0.51 | 0.78 | 0.78 | 0.56 | 0.78 |
| *Panel B: Imputation based on DHS data (not implementable due to incomplete coverage -- locations without DHS data are excluded)* | | | | | | | | | | | |
| State targeting | 37 | 37 | 100% | 37 | 37 | 22,060 | 0.37 | 0.75 | 0.74 | 0.49 | 0.74 |
| LGA targeting | 774 | 631 | 81.52% | 706 | 597 | 19,549 | 0.47 | 0.78 | 0.76 | 0.55 | 0.76 |
| Ward targeting | 8,808 | 1,218 | 13.83% | 2,016 | 464 | 5,968 | 0.54 | 0.83 | 0.79 | 0.66 | 0.79 |
| *Panel C: Imputation based on DHS data (imputing estimates for locations where no DHS data exist)* | | | | | | | | | | | |
| State targeting | 37 | 37 | 100% | 37 | 37 | 22,060 | 0.37 | 0.75 | 0.74 | 0.49 | 0.74 |
| LGA targeting | 774 | 774 | 100% | 706 | 706 | 22,060 | 0.45 | 0.76 | 0.75 | 0.53 | 0.75 |
| Ward targeting | 8,808 | 8,808 | 100% | 2,016 | 2,016 | 22,060 | 0.46 | 0.77 | 0.76 | 0.54 | 0.76 |
| *Panel D: Imputation based on wealth of k-Nearest neighbors in DHS survey* | | | | | | | | | | | |
| *Nearest Neighbor* | - | - | - | - | - | 22,060 | 0.45 | 0.76 | 0.76 | 0.53 | 0.76 |
| *Avg of 5 neighbors* | - | - | - | - | - | 22,060 | 0.46 | 0.76 | 0.76 | 0.52 | 0.76 |

**Table S8 | Targeting simulations in Nigeria. a)** Panel A simulates the performance of an anti-poverty program that geographically targets households in the poorest 2.4km tiles in Nigeria, using the ML estimates of tile wealth. Panels B and C simulate the geographic targeting of households in the poorest states (admin-2), Local Government Areas (LGAs, admin-3), and wards (admin-4), where the most recent DHS survey is used to estimate the average wealth of each administrative region. Panel B ignores households regions with no DHS surveys; Panel C assigns such households the average wealth of the geographic unit closest to the household. Panel D simulates targets poor households where the wealth of a household is inferred based on the average wealth of the households in the DHS cluster physically closest to it. Simulations are evaluated using the 2019 NLSS survey.